\documentclass[pdflatex, sn-mathphys-num]{sn-jnl}% Math and Physical Sciences Numbered Reference Style 

\usepackage{graphicx}%
\usepackage{multirow}%
\usepackage{amsmath,amssymb,amsfonts}%
\usepackage{amsthm}%
\usepackage{mathrsfs}%
\usepackage[title]{appendix}%
\usepackage{xcolor}%
\usepackage{textcomp}%
\usepackage{manyfoot}%
\usepackage{booktabs}%
\usepackage{algorithm}%
\usepackage{algorithmicx}%
\usepackage{algpseudocode}%
\usepackage{listings}%
\usepackage{fullpage}
\usepackage{todonotes}
\usepackage{soul}

\usepackage{lipsum}

\begin{document}

\title[Article Title]{A chemically etched D-band waveguide orthomode transducer for CMB measurements}
% \title[Article Title]{The limit of chemical etching for manufacturing waveguide orthomode transducers above 100\,GHz}
% the limit -> the downside
% the limit -> breaking point?
% the limit -> On the applicability of chemical etching 

%%=============================================================%%
%% GivenName	-> \fnm{Joergen W.}
%% Particle	-> \spfx{van der} -> surname prefix
%% FamilyName	-> \sur{Ploeg}
%% Suffix	-> \sfx{IV}
%% \author*[1,2]{\fnm{Joergen W.} \spfx{van der} \sur{Ploeg} 
%%  \sfx{IV}}\email{iauthor@gmail.com}
%%=============================================================%%

% author list
%\author*[1,2]{\fnm{First} \sur{Author}}\email{elenia.manzan@unimi.it}
%\author[2,3]{\fnm{Second} \sur{Author}}
%\equalcont{These authors contributed equally to this work.}

\author[1,2]{E. Manzan}
\author*[1,2]{A. Mennella}\email{aniello.mennella@fisica.unimi.it}
\author[1,2]{F. Cavaliere}
\author[1,2]{C. Franceschet}
\author[1,2]{S. Mandelli}
\author[1]{F. Montonati}
\author[3,4]{M. Zannoni}
\author[5,6]{P. de Bernardis}
\author[1,2]{M. Bersanelli}
\author[1]{E. Boria}
\author[1]{N. Brancadori}
\author[5,6]{A. Coppolecchia}
\author[3,4]{M. Gervasi}
\author[5,6]{L. Lamagna}
\author[3,4]{A. Limonta}
\author[5,6]{S. Masi}
\author[5,6]{A. Paiella}
\author[3,4]{A. Passerini}
\author[1,2]{F. Pezzotta}
\author[7]{G. Pettinari}
\author[5,6]{F. Piacentini}
\author[8]{E. Tommasi}
\author[2]{D. Vigan\`o}
\author[8]{A. Volpe}

% affiliation
%\affil*[1]{\orgdiv{Department}, \orgname{Organization}, \orgaddress{\street{Street}, \city{City}, \postcode{100190}, \state{State}, \country{Country}}}
\affil*[1]{Università degli studi di Milano, Milano, Italy}
\affil[2]{INFN sezione di Milano, Milano, Italy}

\affil[3]{Università di Milano Bicocca, Milano, Italy}
\affil[4]{INFN sezione di Milano Bicocca, Milano, Italy}

\affil[5]{Università di Roma La Sapienza, Roma, Italy}
\affil[6]{INFN sezione di Roma 1, Roma, Italy}
\affil[7]{CNR/IFN - Consiglio Nazionale delle Ricerche, Roma, Italy}
\affil[8]{ASI - Agenzia Spaziale Italiana, Roma, Italy}

%%==================================%%
%% Sample for unstructured abstract %%
%%==================================%%

\abstract{
This study presents a prototype D-band waveguide orthomode transducer (OMT) fabricated using chemically etched brass platelets. This method offers a fast, cost-effective, and scalable approach for producing such devices above 100\,GHz, making it well-suited for current and future Cosmic Microwave Background polarization experiments, where large focal planes with thousands of receivers are required to detect the faint primordial \textit{B}-modes. Chemical etching has already demonstrated its effectiveness in manufacturing corrugated feedhorn arrays with state-of-the-art performance up to 150\,GHz. Here, we evaluate its applicability to more complex structures, such as OMTs.

We designed a single OMT prototype operating in the 130–170\,GHz range, fabricated by chemically etching 0.15\,mm-thick brass plates, which were then stacked, aligned, and mechanically clamped. Simulations based on metrological measurements of the OMT profile predict {reflected power} below $-$10\,dB, {cross-polarization} better than $-$30\,dB, and transmission around $-$0.5\,dB {in the 140$-$170\,GHz band. Measurements yield transmisison  around $-$1.5/$-$2\,dB and cross-polarization $<-$20\,dB}. Further simulations show that the degradation in the transmission is related to defects and roughness along the etched profile ($\mathrm{RMS}\simeq$3\,$\mu$m), which is a typical and unavoidable effect of chemical etching. The discrepancy in {cross-polarization}, instead, could arise from a  $\sim${1.5}$^{\circ}$ rotation of the polarization angle within the measurement chain.

Our results show that chemical etching is a fast, low-cost, and scalable technique for producing waveguide OMTs with state-of-the-art performance in terms of {reflected power} and {cross-polarization}. However, at frequencies above 100\,GHz {our results show that the mechanical precision limitations of the chemical etching process degrade the transmission coefficient.}
%Therefore, our results show the unsuitability of chemical etching as a manufacturing technique for high-frequency waveguide orthomode transducers.

}


%%\pacs[JEL Classification]{D8, H51}

%%\pacs[MSC Classification]{35A01, 65L10, 65L12, 65L20, 65L70}

\maketitle

%~~~~~~~~~~~~~~~~~~~~~~~~~~~~~~~~~~~~~~~~~~~~~~~~~~~~~~~~~~~~~~~~~~~~~
\section{Introduction}\label{Intro}
%\textcolor{blue}{\lipsum[1-2]} 

In this paper we present a D-band waveguide orthomode transducer (OMT) fabricated with chemically etched brass platelets and evaluate its preformance in the context of current and future Cosmic Microwave Background (CMB) polarization measurements.

The CMB is a relic radiation from the early stage of our Universe that is currently detected as a blackbody spectrum with a brightness peak around 160 GHz \cite{mather1994measurement} and it has proven to be a powerful tool to constrain the cosmological parameters to sub-percent precision \cite{aghanim2020planck}. About 10\% of the CMB intensity is linearly polarized. This fraction can be decomposed into two components, the so-called \textit{E}-modes and \textit{B}-modes \cite{zaldarriaga1997all}. The \textit{E}-modes are due to both scalar (plasma density) fluctuations and tensor (gravitational waves) fluctuations in the primordial plasma, while $B$-modes are generated by tensor fluctuations only. 

While $E$-modes have been characterized with high precision \cite{kovac2002detection, crites2015measurements}, $B$-modes remain a subject of active research. On small angular scales ($\lesssim 1^\circ$) the dominant source of $B$-modes is the gravitational lensing (GL) that CMB photons experience in the proximity of large structures like galaxies or galaxy clusters. On larger scales the $B$-modes are expected to have been generated  by gravitational waves during inflation. While GL $B$-modes have already been measured \cite{wu2019measurement,ade2014measurement}, the ``cosmological'' fraction remains undetected, primarily due to its extreme faintness. {Its detection requires instruments with ultra-high sensitivity with a comparable control of systematic effects and astrophysical foregrounds, which to date are the main limiting factors to the final accuracy}.

{Spurious polarization introduced by non idealities in the detection chain is one of the most critical effects to be controlled and, to this respect, polarizers and OMTs are among the most crucial components.} OMTs are passive polarization dividers that separate incoming radiation into two orthogonal, linearly polarized components {and can be fabricated with microstrips with a planar design or with waveguide structures, symmetric or asymmetric.}

Planar OMTs are generally made of four metal probes on a substrate (silicon) membrane. This design is used in modern CMB experiments above 90\,GHz combined with on-chip filters to provide multi-chroic performance \cite{McMahon_2012,shu2016,anderson2018spt}. They are usually fabricated by deep-ion etching and require additional on-chip power combiners to be coupled to the detectors. {The application of planar designs to OMTs at frequencies above 100\,GHz is still matter of technological research and development, as shown by recent works by Valente and Navarrini in the W-band \cite{valente2024} and Lu et al in the D-band \cite{lu2025}}. Typically, planar OMTs display {reflected power (RP)} $<-15$\,dB, which is mainly limited by power leakage in the coupling to the antennas. The {cross-polarization} is generally $<-20$\,dB, limited by crosstalks between the probes and the detectors within the transmission line.

Waveguide OMTs, symmetric or asymmetric, notably perform better than planar ones. Symmetric OMTs display wider bandwidth (typically $30\%-40\%$) and good electromagnetic performance ({RP} $<-20$\,dB, {cross-polarization} $<-40$\,dB), though at the expense of a mechanically complex design. Several prototypes have been fabricated using milling \cite{narayanan2003full,navarrini2006test,Navarrini_2011_test,pisano2007broadband}, electroforming \cite{asayama2009development} or silicon micro-machining \cite{gomez2018compact} up to $500$\,GHz. 

Asymmetric waveguide OMTs, like the one presented in this work, feature a simpler mechanical designs while achieving electromagnetic performance comparable to that of symmetric OMTs at the cost of a narrower bandwidth (less than $20\%$). They can also be designed for dual-band operation \cite{el2018dual,leal2013micromachined}. Several prototypes have been fabricated by milling \cite{dunning2009simple}, micro-milling \cite{reck2013600}, electroforming \cite{d2009planck}, {wire-erosion \cite{virone2014}}, reaching frequencies up to 600\,GHz. Although all the aforementioned manufacturing techniques provide tight mechanical tolerance ($\simeq 30\ \mu$m), they are generally either expensive or time-consuming, and therefore not ideal for producing large arrays.

{The platelet technique, introduced by Haas in 1993 \cite{haas1993,haas1995}, represents an alternative approach to reduce fabrication costs and improve scalability. It requires drilling holes into metal plates that are subsequently stacked and clamped or bonded to form the desired cavity structure and it has been widely applied to fabricate corrugated feedhorns (see, for example, Del Torto et al \cite{deltorto2011}, Simon et al \cite{simon2018}, Mandelli et al \cite{mandelli2021}, Cavaliere et al \cite{cavaliere_2022}).}

{The application of the platelet technique to the platelet technique to OMTs, however, is much less documented in the literature, especially at high frequencies \cite{cano2018,peverini2022}.} Our work represents the first waveguide OMT {in the D-band} manufactured {in platelets}. {Furthermore we used chemical etching, which is  a very low-cost manufacturing technique that consists of the controlled erosion of metal plates through an acid solution.} Since chemical etching permits the drilling of several plates simultaneously with a mechanical tolerance of $50/30\ \mu$m, it can be applied to produce large arrays with hundreds of elements. Indeed, it has already proven successful in fabricating corrugated feed-horn arrays with performance comparable to the state-of-the-art up to 150\,GHz \cite{Mandelli_2021,cavaliere_2022}.

{Before producing a full array, as a first step we realized a single asymmetric D-band ($130-170$\,GHz) OMT that we coupled to the central antenna of the 37-element corrugated feed-horns array described in Mandelli et al} \cite{Mandelli_2021} on one side and to a pair of kinetic inductance detectors (KIDs) on the other. {Our goal was to obtain performance typical of asymmetric structure as, for example, the OMTs used by the Planck-LFI instrument \cite{d2009planck}, which feature {reflected power} between $-$10\,dB and $-$25\,dB plus cross-polarization and isolation of the order of $-$40\,dB over 20\% of the band.}

In Section \ref{Design} we detail the electromagnetic and mechanical design of the OMT prototype. In Section \ref{Manuf} we show the mechanical measurements of the thin etched plates assessing the achieved manufacturing tolerance, along with the simulated frequency response based on the measured OMT profile. In Section \ref{EMmeas} we present the experimental setup and discuss the measured frequency response. In Section \ref{Concl} we summarize the main results and discuss possible improvements and future developments.

\section{OMT Design}\label{Design}

    \subsection{Electromagnetic design}\label{EMdesign}
    
\begin{figure}[ht]
        \begin{center}
            \includegraphics[width=9cm]{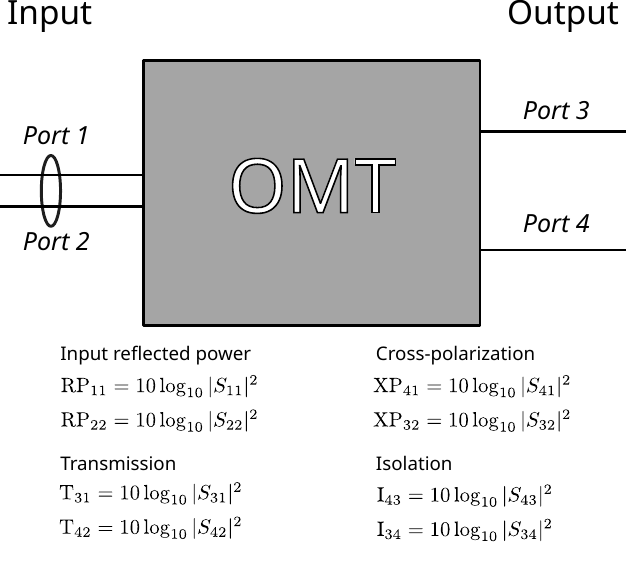}  
            \caption{Block schematic of an OMT with the definition of the main electromagnetic parameters. {Ports 1 and 2 appear grouped by the black ellipse as they correspond to a single physical port, the input circular waveguide connected to the feed horn.}}
            \label{fig:omt_schematic}
        \end{center}
    \end{figure}

    {Figure \ref{fig:omt_schematic} shows a block schematic of an OMT. {The left ports 1 and 2 represent the orthogonal input polarizations. In the figure they appear to be grouped by the black ellipse as they correspond to a single physical port, the input circular waveguide connected to the feed horn.} Ports 3 and 4 on the right indicate the two rectangular waveguides for the output orthogonally polarized signals. Below the block we define the input {reflected power}, transmission, cross-polarization and isolation in terms of the scattering matrix parameters, $S_{i,j}$.}
    
    {Our OMT is similar to that described in D'Arcangelo et al \cite{darcangelo_2009} and consists of a main and a side arm (see top-left of Fig.~\ref{fig:em_designs})}. A circular-to-square waveguide transition is followed by a T-junction polarization divider that splits the incoming radiation into two orthogonal, linearly polarized components, each of which propagates in one of the two arms. The main arm is completed with a 3-step transition to the output, rectangular waveguide. The side arm has a $90^{\circ}$ H-bend that rotates the radiation parallel to the main axis, towards a step transition to the output, rectangular waveguide, such that the two output ports lie on the same plane. The intermediate square waveguide limits the bandwidth to around $25\%$, because the cut-off frequency of the first mode is $\sqrt{2}$ times that of the fundamental mode, rather than twice the value as in rectangular waveguides.

    We modeled and optimized the OMT design using the transient, time domain, solver of the CST Microwave Studio Suite, an electromagnetic simulator based on the Finite Integration Technique (FIT)\footnote{https://www.3ds.com/products/simulia/cst-studio-suite}.

        \subsubsection{Platelet design} \label{Nominal_OMT_design}

\begin{figure}[ht]
		\hfil\includegraphics[width=6cm]{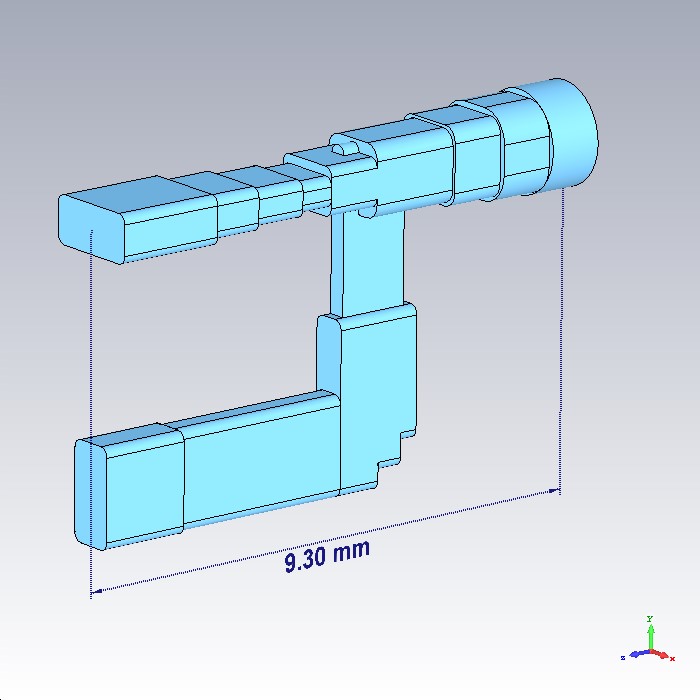}
		\includegraphics[width=7.5cm]{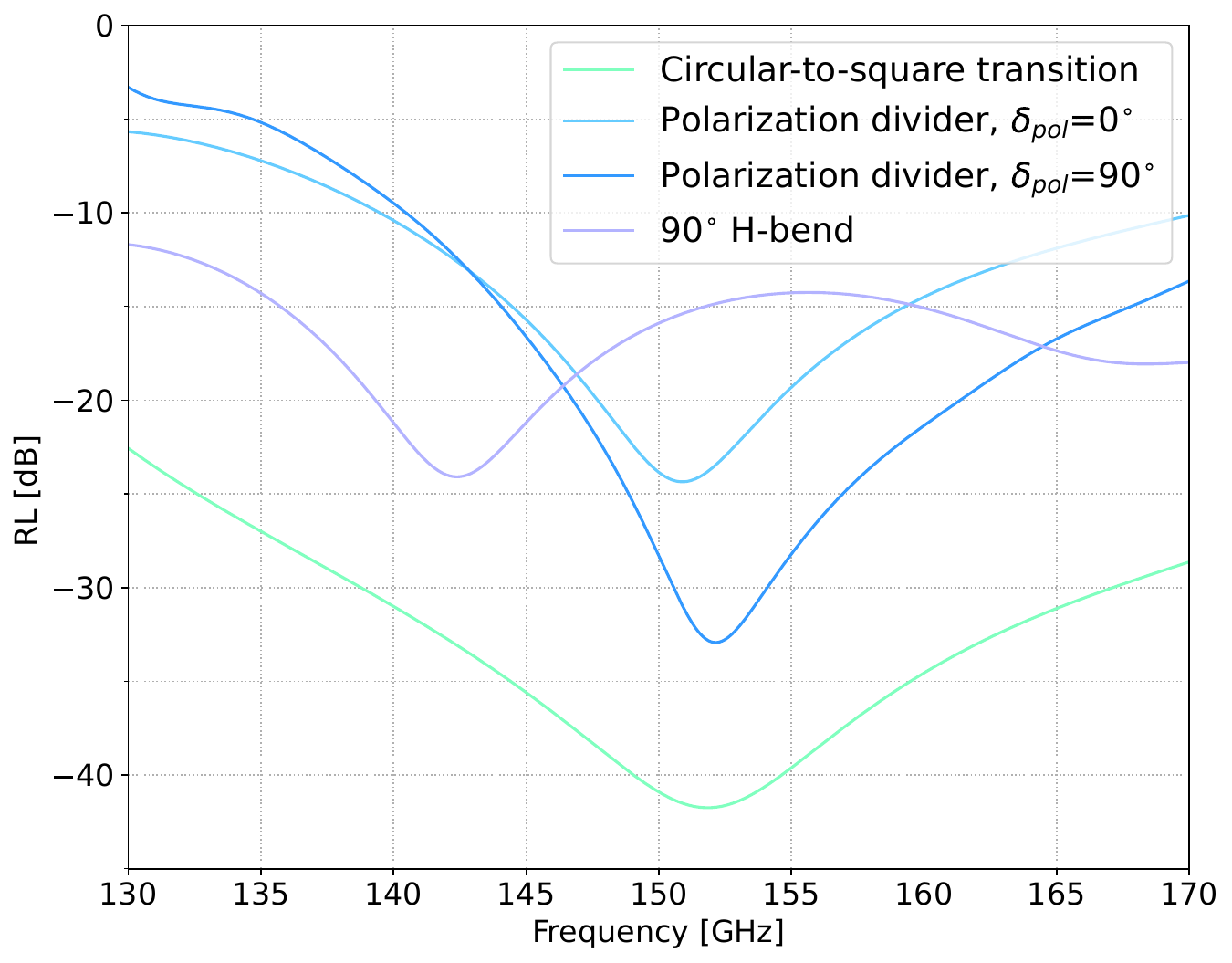}  \newline 
		\centering
        	\includegraphics[width=6cm]{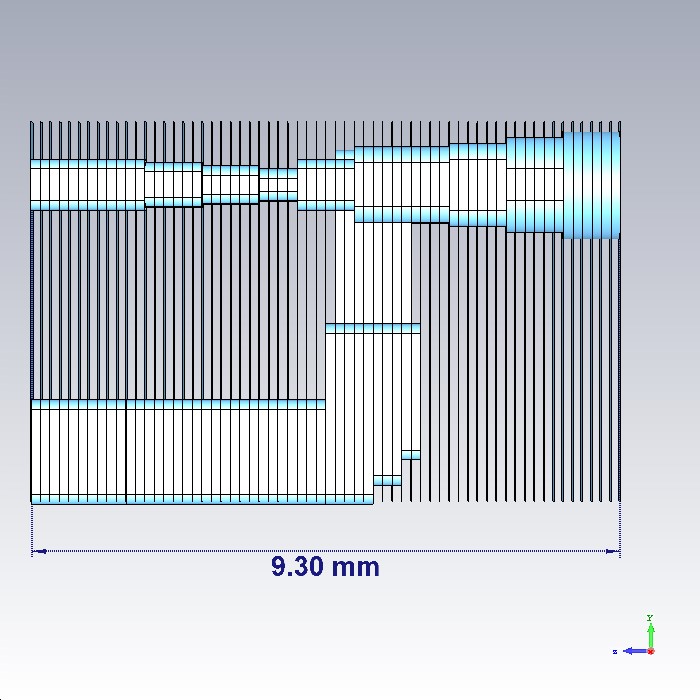}  
        	\includegraphics[width=7.5cm]{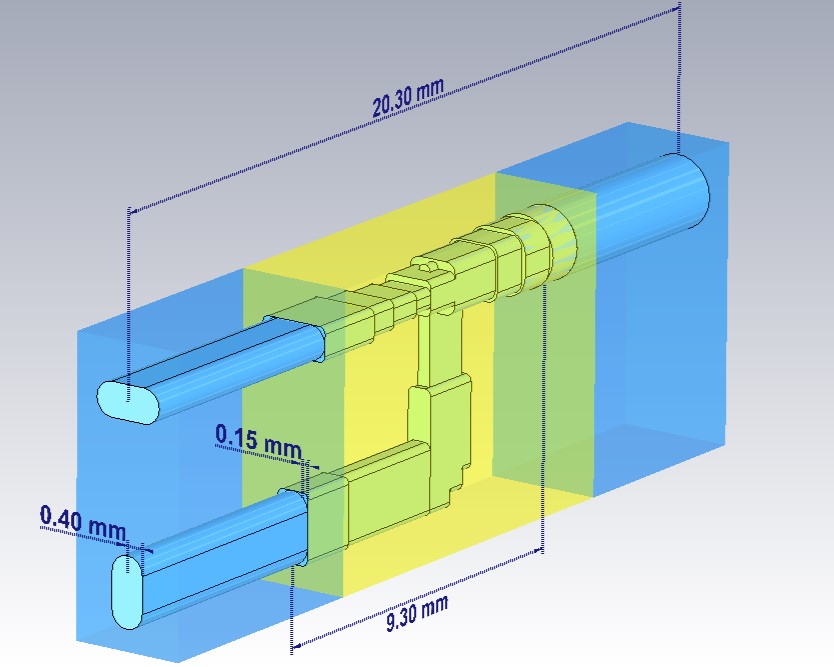} 
        	\caption{\textit{Top-left}: OMT design. The image shows the inner structure, in light blue, where radiation propagates. \textit{Top-right}: optimized performance of the OMT main components. We show the polarization splitter response in the case of horizontal ($\delta_{\mathrm{pol}}=0^{\circ}$) and vertical ($\delta_{\mathrm{pol}}=90^{\circ}$) incoming polarization.\textit{Bottom-left}: image showing how the OMT can be subdivided into thin slices perpendicular to its axis, represented by the vertical lines, which permits platelet manufacturing. \textit{Bottom-right}: complete, nominal OMT design. The OMT is packed between a circular input waveguide and two oval output waveguides. The colors represent the materials used: the OMT is made of brass (yellow) and the closure plates of aluminum (blue). This is also called \textit{trade-off} design.}
     		\label{fig:em_designs}
    \end{figure}

        We designed the OMT considering the mechanical constraints detailed in Table \ref{OMT_constraints}, specifically: 
        \begin{enumerate}
            \item the plate thickness of 0.15\,mm was required to ensure a mechanical tolerance of $\pm 30\ \mu$m obtained by chemical etching. This implied that the thickness of each OMT component had to be a multiple of this value, as shown in the bottom-left panel of Fig.~\ref{fig:em_designs}; 
            \item the circular waveguide diameter and the center-to-center distance between the two output ports were fixed to 1.7\,mm and 4.22\,mm, respectively, to enable coupling between the OMT, a pre-existing antenna, and two KID detectors on the focal plane; 
            \item the dimensions of the output waveguides conform to the standard rectangular waveguide dimensions for the D-band. 
        \end{enumerate}

        Additionally, the waveguides have rounded edges, with a radius varying from 0.5\,mm to 0.15\,mm, since chemical etching does not allow to obtain sharp corners.
        
        We designed each component individually, through a parametric optimization process that fine-tuned the cavity dimensions to meet the performance requirements. In our case, we minimized the {reflected power} across the $130-170$\,GHz band, with a $-20$\,dB requirement at the center of the band. We further optimized the design to ensure a {cross-polarization} below $-50$\,dB. The top-right panel of Fig.~\ref{fig:em_designs} displays the optimized performance of the input transition, the T-junction, and the H-bend.

        \begin{table}[ht]
        \caption{Mechanical constraints for the OMT design}\label{OMT_constraints}
        \begin{tabular}{@{}ll@{}}
        \toprule
        Parameter & Requirement \\
        \midrule
        Plate thickness & 0.15\,mm \\
        Circular waveguide diameter & 1.7\,mm  \\
        Rectangular waveguide major axis & 1.65\,mm \\
        Rectangular waveguide minor axis & 0.8\,mm  \\
        Center-to-center rectangular waveguides distance & 4.221\,mm \\
        Rounded corners radius & $0.15$$-$$0.5$\,mm \\
        %Main arm length & 9.3\,mm  \\
        \botrule
        \end{tabular}
        \end{table}

        \subsubsection{Closure plates} \label{Top_Bottom_design}
        We designed our prototype to be mechanically clamped, a process that eliminates the need for glue or bonding, allowing for disassembly and replacement of individual plates, if needed. The platelets are stacked between 5\,mm thick top plate (housing the 1.7\,mm circular waveguide) and three 2\,mm thick bottom plates (containing the output rectangular waveguides), which provide the necessary structural stiffness. We designed the closure plates to be manufactured through CNC milling since they are too thick for chemically etching.

        The radius of curvature of the output waveguides is 0.4\,mm, which does not match that of the last platelet (0.15\,mm). {Our simulation showed that this mismatch does not affect the performance significantly so we accepted this curvature radius as it simplifies the fabrication process.} As a result, the output waveguides have an oval shape, as shown in the bottom-right panel of Fig.~\ref{fig:em_designs}. 

        \subsubsection{Nominal frequency response} \label{Nominal_response}
\begin{figure}[ht]
		\hfil\includegraphics[width=16cm]{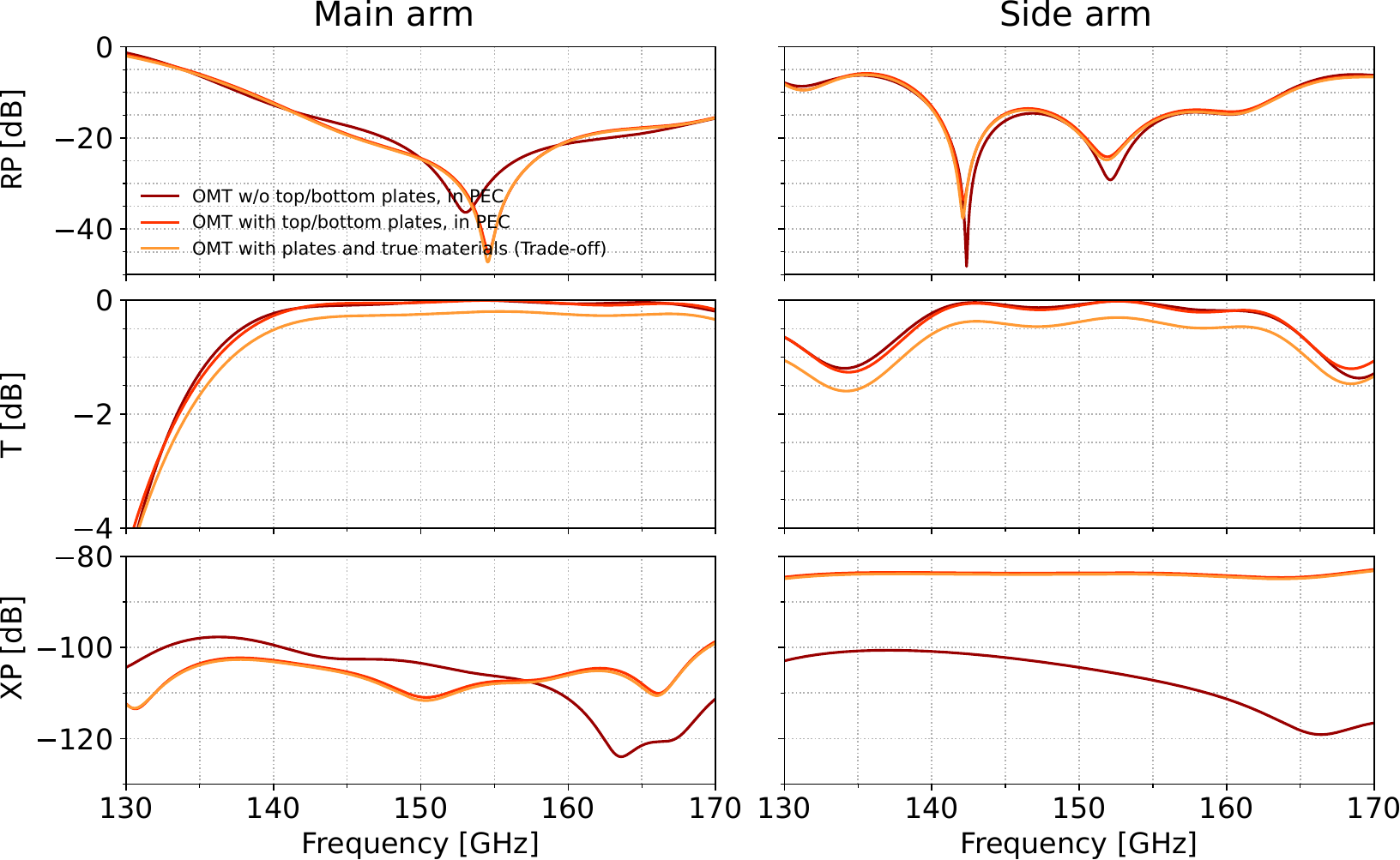}
        	\caption{Simulated, nominal scattering parameters. The dark-red curves represent the nominal OMT without closure plates, assuming PEC material. The OMT completed with the closure plates is plotted in red, assuming PEC. The transition from the rectangular to the oval output does not significantly alter the performance. The same OMT, but assuming the real manufactured materials, brass, and aluminum, is plotted in orange. The transmission due to the finite conductivity of the materials is around $-0.5$\,dB.}
     		\label{fig:sim_em_nominal}
    \end{figure}

        In the nominal case of a perfect electric conductor (PEC) material and a perfect geometrical profile without the closure plates we obtain the simulated frequency response displayed in Fig.~\ref{fig:sim_em_nominal}. For each arm, we report the scattering parameters related to {reflected power} ({RP}), transmission (T), {coss-polarization (XP) and isolation (I)} coefficients. {Referring to the schematic of Fig.~\ref{fig:omt_schematic}, for the main arm we have $\text{{RP}}(\text{main\,arm})\equiv \text{{RP}}_{11}$, $\text{T}(\text{main\,arm})\equiv \text{T}_{31}$, and $\text{XP}(\text{main\, arm})\equiv \text{XP}_{41}$; similarly, for the side arm we have $\text{{RP}}(\text{side\,arm})\equiv \text{{RP}}_{22}$, $\text{T}(\text{side\,arm})\equiv \text{T}_{42}$, and $\text{XP}(\text{side\, arm})\equiv \text{XP}_{32}$.}
        
        The nominal case features $\text{{RP}}<-10$\,dB over most of the $130$$-$$170$\,GHz range and $<-20$\,dB at the center of the band. The {cross-polarization} is less than $-80$\,dB and transmission is close to $0$\,dB. Adding the closure plates does not significantly alter the expected performance. Finally, we simulated the OMT response assuming the true materials, a combination of brass and aluminum, as further discussed in Section \ref{MECdesign}: due to the finite conductivity of the materials, the more realistic expected transmission is around $-0.5$\,dB. We call this case \textit{trade-off} design and we will assume this configuration in the rest of the paper. {The name ``trade-off'' was chosen because this case was a compromise between the ideal performance and the feasibility constraints.}

    \subsection{Mechanical design}\label{MECdesign}

\begin{figure}[ht]
		\centering
		\includegraphics[width=7cm]{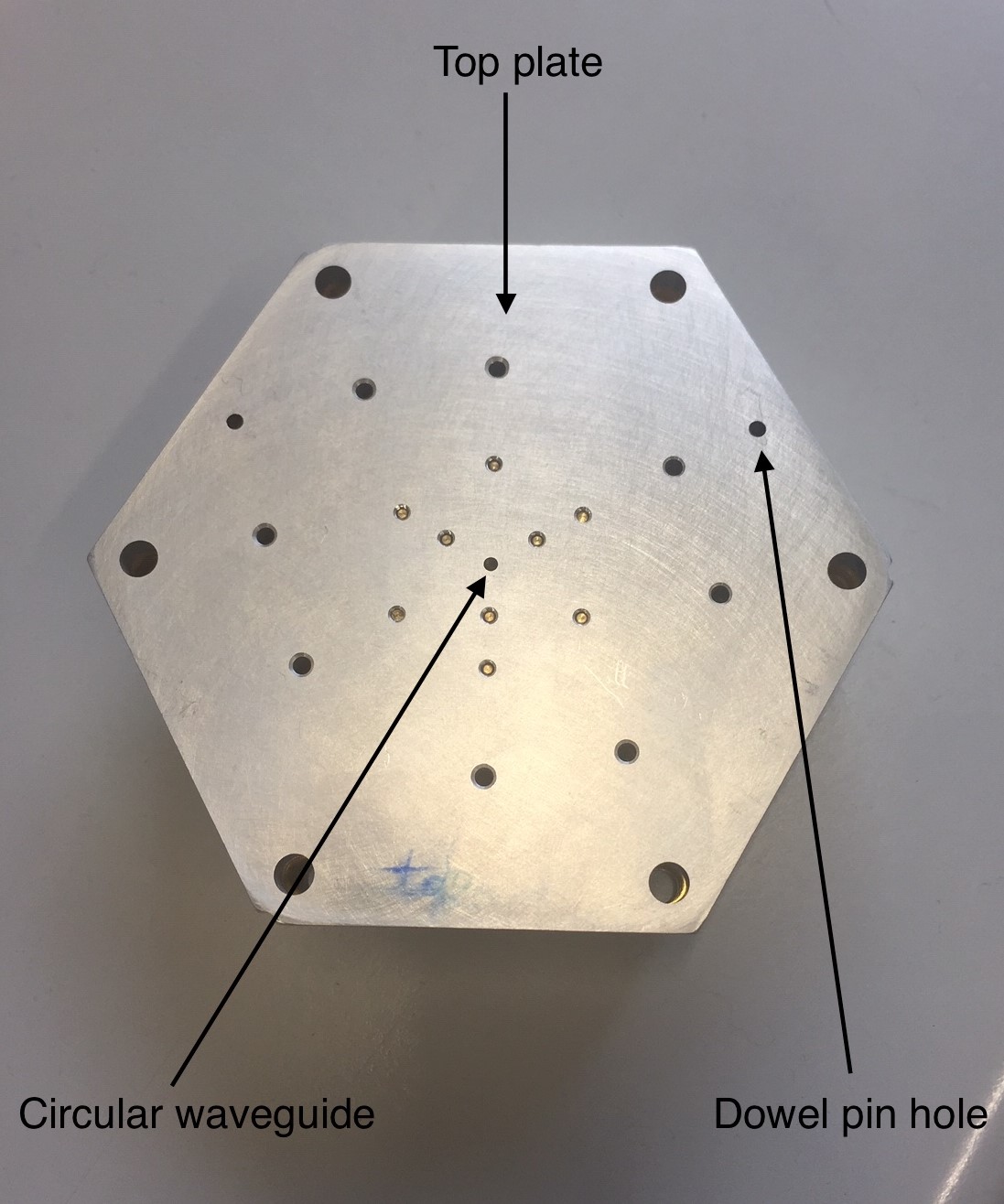}
        	\includegraphics[width=7cm]{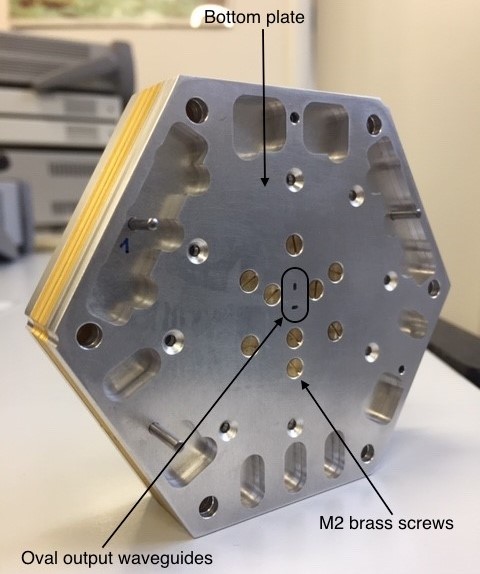} \newline
        	\includegraphics[width=5.45cm]{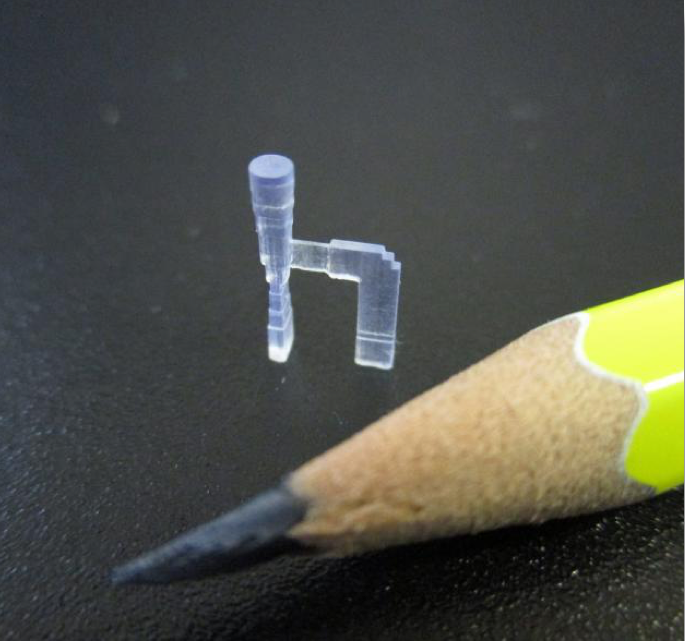} 
        	%\centering
        	\includegraphics[width=5cm]{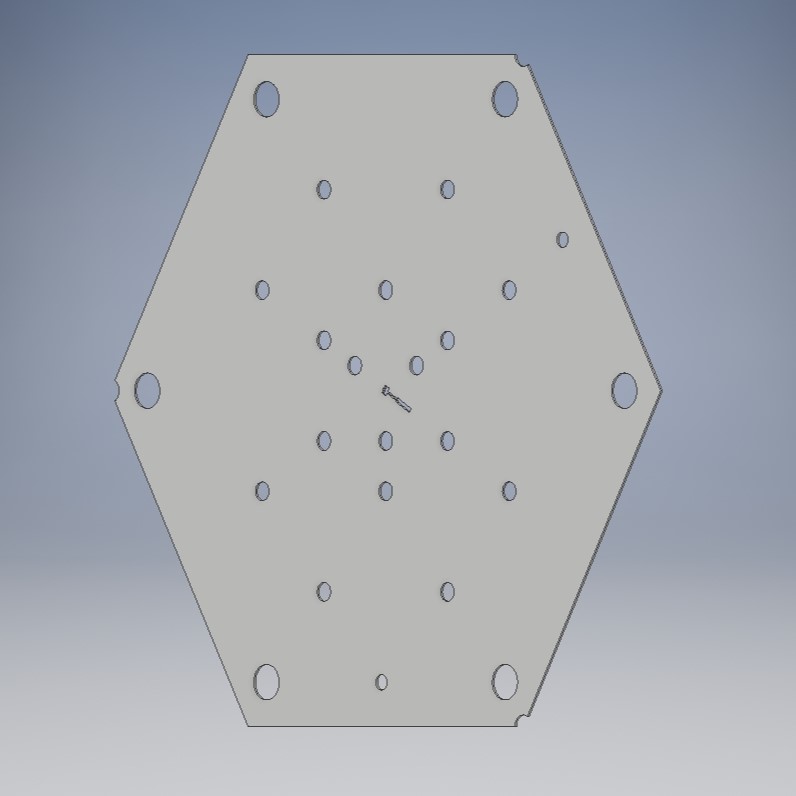} 
        	\includegraphics[width=5cm]{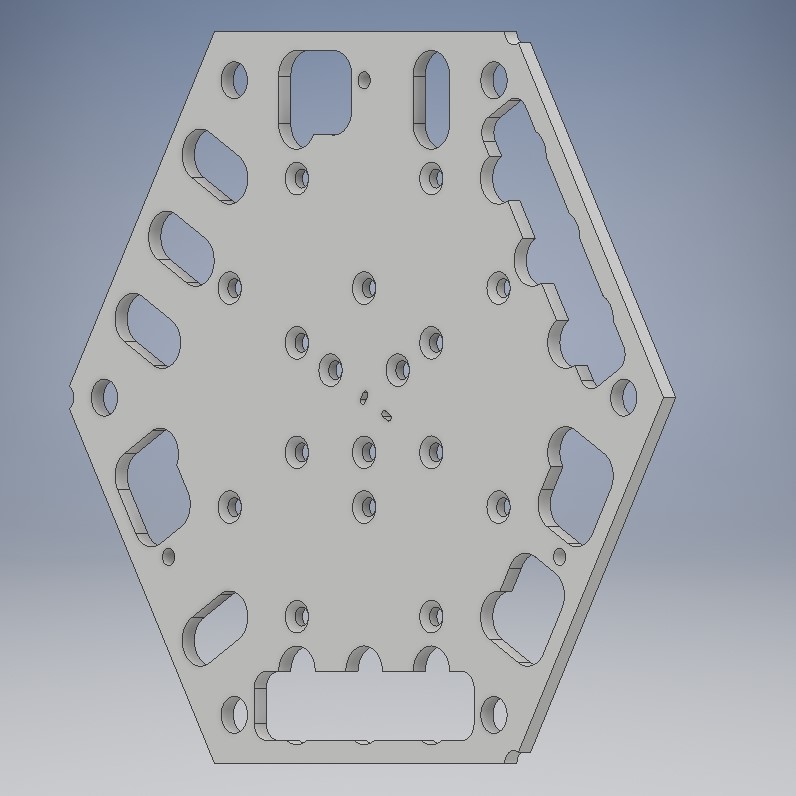} 
        	\caption{\textit{Top row}: Front and rear view of the OMT. \textit{Bottom row}: 3D-printed mockup of the inside of the OMT (left), design of a thin, intermediate plate (center), and one of the bottom plates with the custom outer trench (right).}
     		\label{fig:mec_design}
	\end{figure}

    The top row of Fig.~\ref{fig:mec_design} displays the prototype front and rear views, whereas the bottom-left panel shows a 3D-printed mockup of the OMT profile inside the metal structure.
    
    We designed our OMT to be coupled to the central antenna of a preexisting feed-horn array, described in Mandelli et al \cite{Mandelli_2021}, to illuminate a KID focal plane. The feed-horn system consists of 37 corrugated feed-horns distributed with a center-to-center distance of 11\,mm on a hexagonal footprint of 49\,mm side. Our prototype shares the same hexagonal geometry and consists of 62 thin brass plates of 0.15\,mm thickness overlapped between a 5\,mm thick top plate and a 2\,mm bottom plate, both made of aluminum. {Each plates includes two holes designed for 2\,mm rectified steel dowel pins, for mechanical alignment plus 17 M2 and six M4 threaded holes for brass screws for clamping}. The CAD design of a thin plate is shown in the bottom-center panel of Fig.~\ref{fig:mec_design}.
    
    The prototype is completed by two additional, twin bottom plates of 2\,mm thickness, hosting the same oval output waveguides and screw holes of the first bottom plate, but with a custom outer trench to interface with the focal plane. The CAD design of this last bottom plate is shown in the bottom-right panel of Fig.~\ref{fig:mec_design}. Since the trench is in line with the OMT dowel pins, we inserted three additional pin holes in each of the three bottom plates to align them together. We used the last, 2\,mm thick, interface plate to host the screw heads and we used the thicker top plate to tighten the screws, as well as place the dowel pins and guarantee their orthogonality to the plates during the assembly. 

    The material and thickness of the thin, intermediate plates have been determined by manufacturing constraints since the chemical erosion of the small structures that constitute our platelet design is guaranteed with a mechanical tolerance of $\pm30\ \mu$m only for 0.15\,mm brass plates.
    
%     \todo[size=tiny]{I need feedback on the following part from people who know better than me. Is this part convincing?}
    
    In the context of our exploratory study, we chose brass over more lightweight and conductive materials, like aluminum, because of its compatibility with chemical etching. The expected OMT transmission is not significantly different from that of an aluminum counterpart, and could be further improved by silver plating if necessary. Additionally, weight constraints were not critical in our case. However, weight could become a limiting factor in space applications.
    
    %{The screws material was instead selected for its convenient machinability. Brass produces much less chips during milling or turning and is therefore preferable to aluminum for manufacturing small-diameters screws sush as M2 or M4.}
    
    {A possible downside in the use of brass is the linear expansion coefficient ($19 \times 10^{-6}\ ^{\circ}$C$^{-1}$) being slightly different from that of aluminum ($24 \times 10^{-6}\ ^{\circ}$C$^{-1}$), constituting the rest of the receiver. However, the limited dimensions of the devices imply that the difference in contraction between one material and the other at the operating temperature ($\sim 0.1 $\,K) is of the order of the mechanical tolerance itself. {Therefore, any thermally induced misalignment can be considered acceptable.}}
    
%     	{The use of brass screws was mainly driven by project schedule constraints. Although brass does not provide a perfect match to aluminum in the coefficient of thermal expansion, (CTE$_\text{Al}\sim23\,\mu\text{m}\text{K}^{-1}$, CTE$_\text{Brass}\sim21\,\mu\text{m}\text{K}^{-1}$ for brass), the differential contraction between the screws and the block of aluminum plates during operations at 4\,K is of about 10\,$\mu$m, that was considered acceptable in the context of a demonstration prototype.}

    {From a clamping perspective, differential thermal contraction, could theoretically reduce clamping torque applied at room temperature. Given the complex geometry of the clamping system, a full thermomechanical analysis would be required to precisely quantify this effect. In this work, since our functional assessment was performed at room temperature, we decided anyway to adopt different materials, also due to in-house availability constraints. Nevertheless, for cryogenic testing, we recommend the use of matching materials.}
    
    Finally, we note that for thicker plates or more relaxed tolerance constraints, as could be the case for lower-frequency devices, chemical etching could be used on aluminum plates, too. 
    
\section{Manufacturing}\label{Manuf}
%manufacturing process
The thin plates have been chemically etched at Lasertech SRL in Milan, whereas the closure plates have been fabricated at the workshop of our Physics Department. The etching process involves stamping the mechanical design onto a photo-sensitive master layered on a thin metal sheet. This sheet is then exposed to UV radiation, imprinting the design, and subsequently immersed in an acid bath, where the unimpressed material is progressively eroded. The fabrication of the closure plates consists of drilling circular or oval holes by means of a computer numerically controlled (CNC) milling machine. 

    \subsection{Metrological measurements for quality assessment}\label{Metrology}

\begin{figure}[ht]
        \includegraphics[width=8cm]{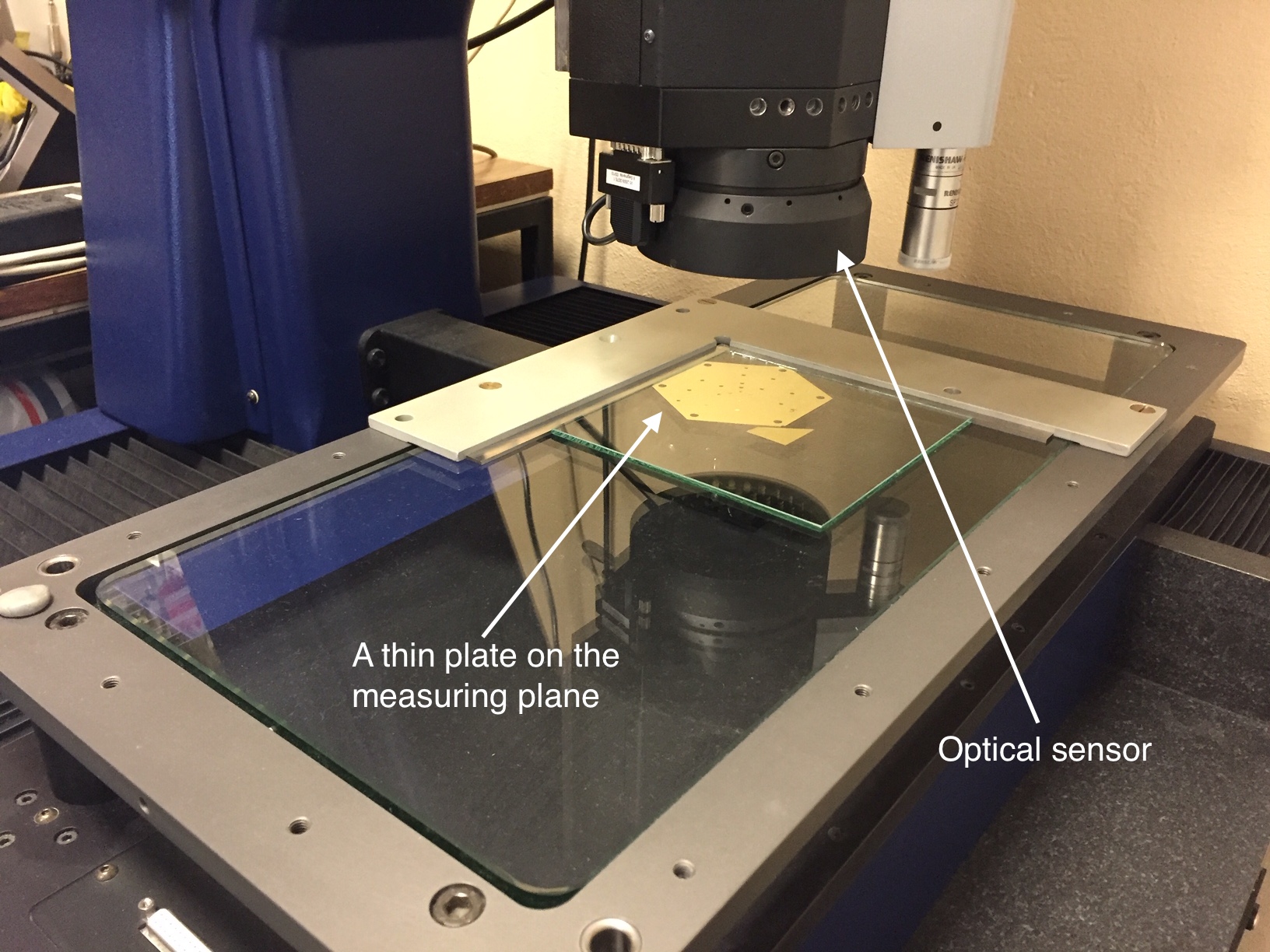}
        \includegraphics[width=8cm]{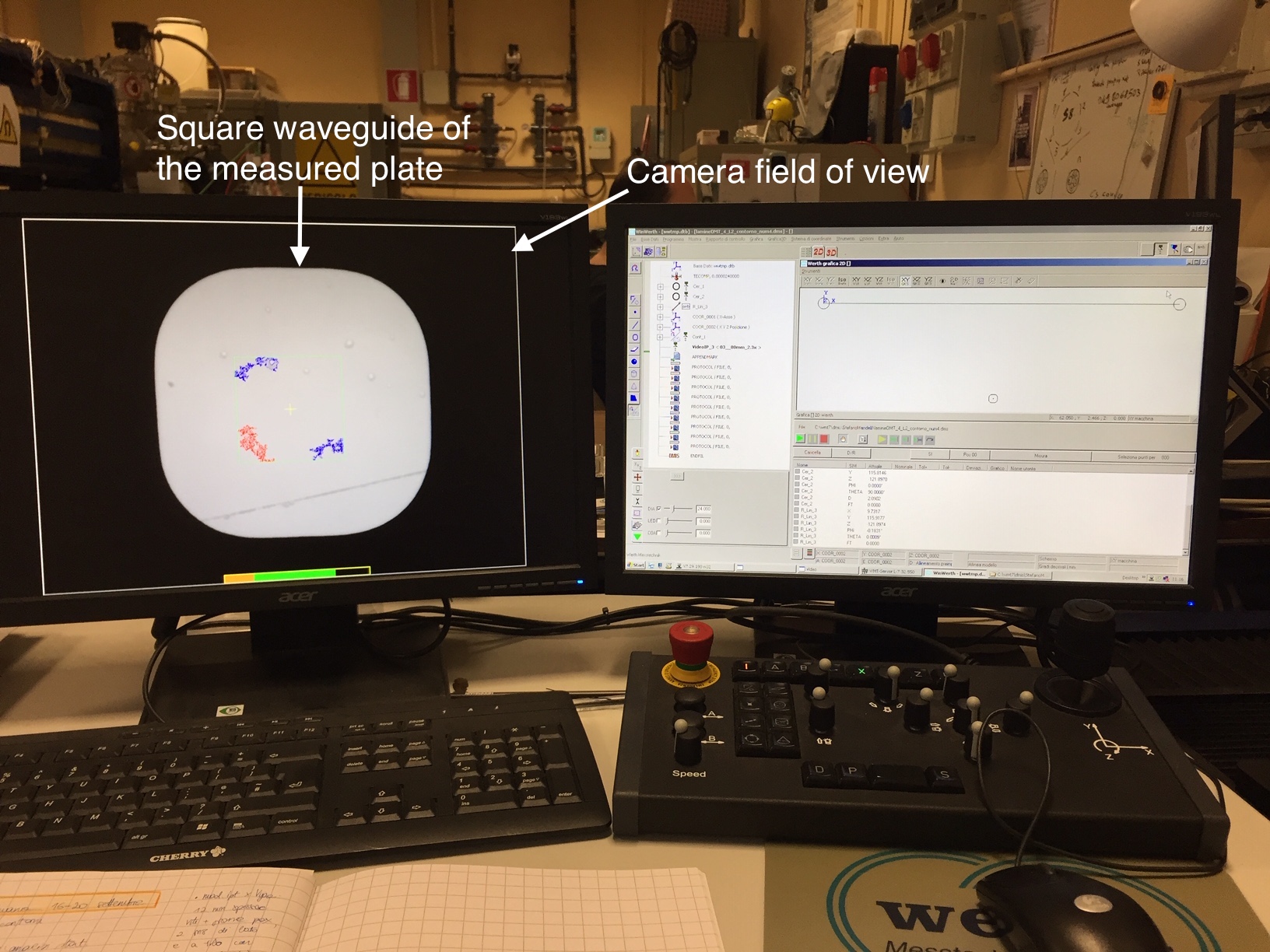}
        \caption{\textit{Left panel}: Werth ScopeCheck 200 metrological machine measuring one of the thin plates. \textit{Right panel}: Machine control monitors and console.}
        \label{fig:werth_image}
	\end{figure}

    %metrological measurements to asses manufacturing accuracy
	We performed metrological measurements of all the plates using our Werth ScopeCheck 200 machine, shown in Fig.~\ref{fig:werth_image}, to asses the achieved fabrication precision against the maximum achievable tolerance. The mechanical tolerance of the chemical etching is $\pm$0.03\,mm on 0.15\,mm thick brass plates, whereas the CNC milling precision is $\pm$0.03\,mm. 
	
	For each pin hole and OMT profile, we measured its dimensions (diameter or rectangular axes) and the $x$, and $y$ center coordinates, comparing them to their nominal value. In the case of circular profiles, we also measured the deviation from circularity (form tolerance, FT\footnote{The form tolerance of a round shape is the width of the annular ring containing the shape.}). FT measurements are sensitive to any shavings or impurities along the measured profile but, according to our experience in fabricating chemically etched feedhorn arrays, a FT $< 10\,\mu$m indicates no significant deviation from circularity.
 
    The measurement of the rectangular profiles could not be completely automated and this made it more prone to systematic errors introduced by the operator. We checked these measurements for each platelet by comparing the deviations of the alignment pin center positions with the deviations of the rectangular profiles and verified that both were consistently less than about 10\,$\mu$m.

\begin{figure}[ht]
 		\includegraphics[width=16cm]{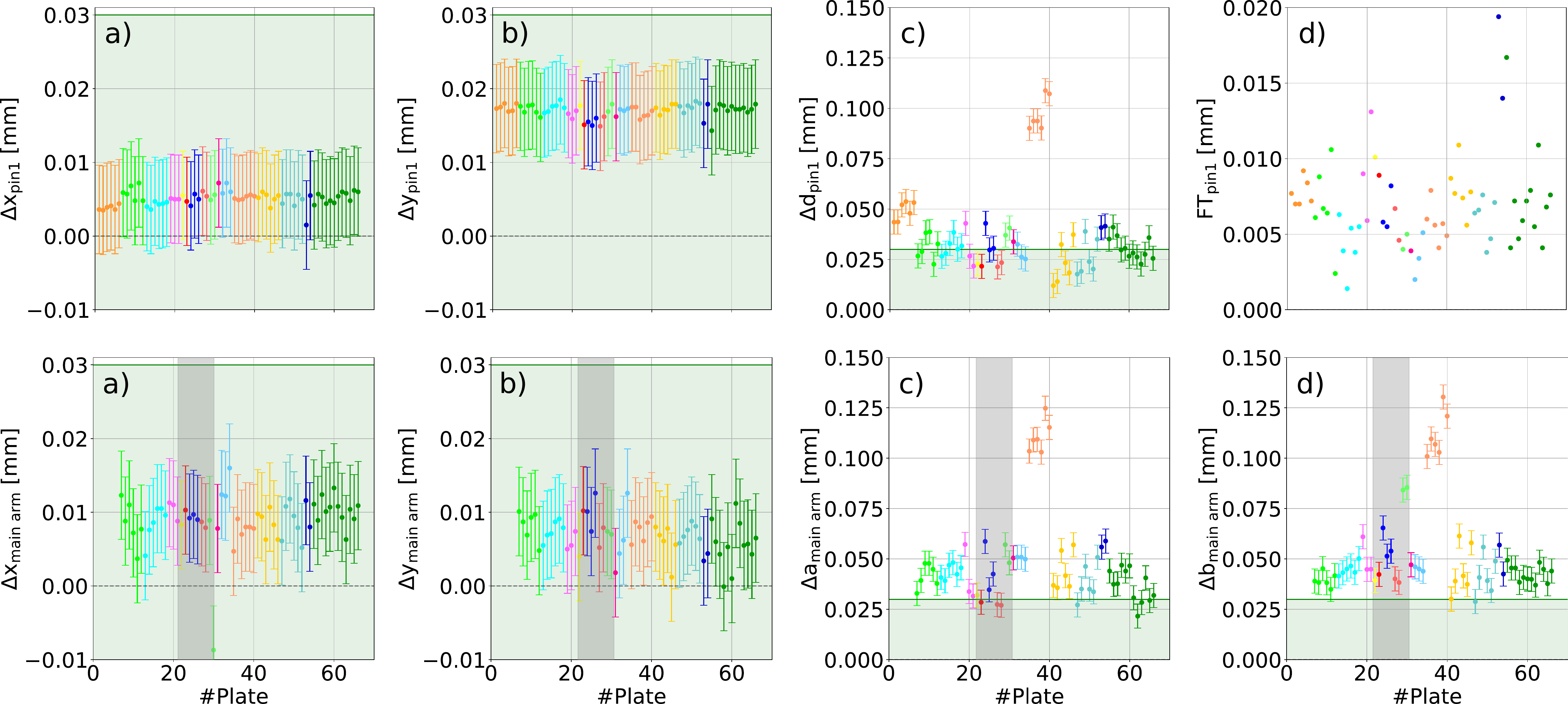} 
	        \caption{\textit{Top row}: Metrological measurements of one of the two pin holes. Deviation, $\Delta$, between measured and nominal: $x$ (a) and $y$ (b) center coordinates, diameter $d$ (c), and measured FT (d), which indicates a deviation from circularity. \textit{Bottom row}: Metrological measurements of the main arm. Deviation between measured and nominal: $x$ (a) and $y$ (b) center coordinates, major waveguide axis $a$ (c), and minor waveguide axis $b$ (d). Measurements of the same color belong to the same section of the OMT (e.g. \textit{dark green} for the rectangular waveguide plates). The green area highlights the expected manufacturing tolerance, $\pm30\ \mu$m, while the grey area highlights the plates forming the polarization divider, which were difficult to measure and could be overestimated. Note that, since we are showing the square and rectangular waveguide of the main arm, the first six plates containing the circular waveguide are missing.
	        }
     		\label{fig:metrol_meas}
	\end{figure}

	Figure~\ref{fig:metrol_meas} shows the measurements of one of the two pin holes and of the OMT main arm. Similar results were obtained for the other pin and the OMT side arm, but are not shown here to avoid redundancy. %Measurements of the same colour belong to the same section of the OMT (e.g. \textit{yellow} for the circular waveguide plates and \textit{dark green} for the rectangular output waveguides plates). The green area highlights the expected manufacturing tolerance, while the grey area in the bottom-right section of Fig.~\ref{fig:wg3_meas} highlights the set of plates that constitute the polarization divider.
    The $\pm6\ \mu$m error bar is the metrological machine intrinsic uncertainty.

	The $x$ and $y$ center coordinates comply with the mechanical tolerance and show a systematic shift towards positive deviations: the measurements are closely scattered around an average value of $\Delta x=(5\pm6)\ \mu$m, $\Delta y=(17\pm6)\ \mu$m for the pins and $\Delta x=(9\pm6)\ \mu$m, $\Delta y=(7\pm6)\ \mu$m for the OMT. The pin diameters also comply with the mechanical tolerance, with a few outliers, and the FT is on average below $10\ \mu$m, showing no significant deviation from circularity.
 
     Most of the major and minor axes are slightly out of spec. Since the results are consistent with the pin measurement within 10--20\,$\mu$m, we exclude the possibility that this discrepancy arises from an error in our measurement strategy. Instead, we attribute it to the erosion rate: by the time the acid had fully eroded the 2 mm pinhole, it had already over-eroded the OMT profile, whose nominal dimensions are slightly smaller than those of the pin.

    The grey areas in the bottom panels of Fig.~\ref{fig:metrol_meas} represent measurements of the minor axis of the polarization divider. This measurement was particularly challenging due to the presence of rounded corners within the T-junction, which may have introduced systematic uncertainties not accounted for by the error bars shown in the graphs.

    %On the systematic effects
	Overall, all the measurements are systematically larger than their nominal value, especially those of a group of 6 plates (1 sheet) after the polarization division (orange points). This over-erosion results from the limited control of the erosion process and has been noticed in our previous work on feed-horn arrays \cite{Mandelli_2021,cavaliere_2022}. {We discuss the effects of these systematic deviations on the OMT performance in Section~\ref{Sim_with_systematics}}. Moreover, the over-erosion measurements allow us to compensate for this systematic effect in future prototypes by properly rescaling the dimensions in the mechanical design.

	The measurements of the closure plates comply with the milling mechanical precision and are omitted for simplicity.

	The array was manually aligned using two orthogonal grinded surfaces and rectified steel dowel pins. We tightened the OMT using M2 brass screws and qualitatively checked the alignment of the first plates with our Werth machine. We used the seven inner screws to clamp the OMT, so that the remaining M2 screws can be used to couple the OMT to the feed-horn array and the M4 screws to couple the overall system to its focal plane.
    
    \subsection{Simulations including fabrication systematic effects}\label{Sim_with_systematics}
	We performed electromagnetic simulations on a model OMT based on the measured profile to assess the performance reduction caused by imperfections and misalignments introduced during the assembly. We explain the strategy that allowed us to reconstruct and simulate the measured profile in Section~\ref{metrol_point_acq} and show the results of the electromagnetic simulations in Section~\ref{Sim_of_metrol_OMT}. 

        \subsubsection{Metrological point-cloud acquisition} \label{metrol_point_acq}

\begin{figure}[ht]
 		\centering\includegraphics[width=15cm]{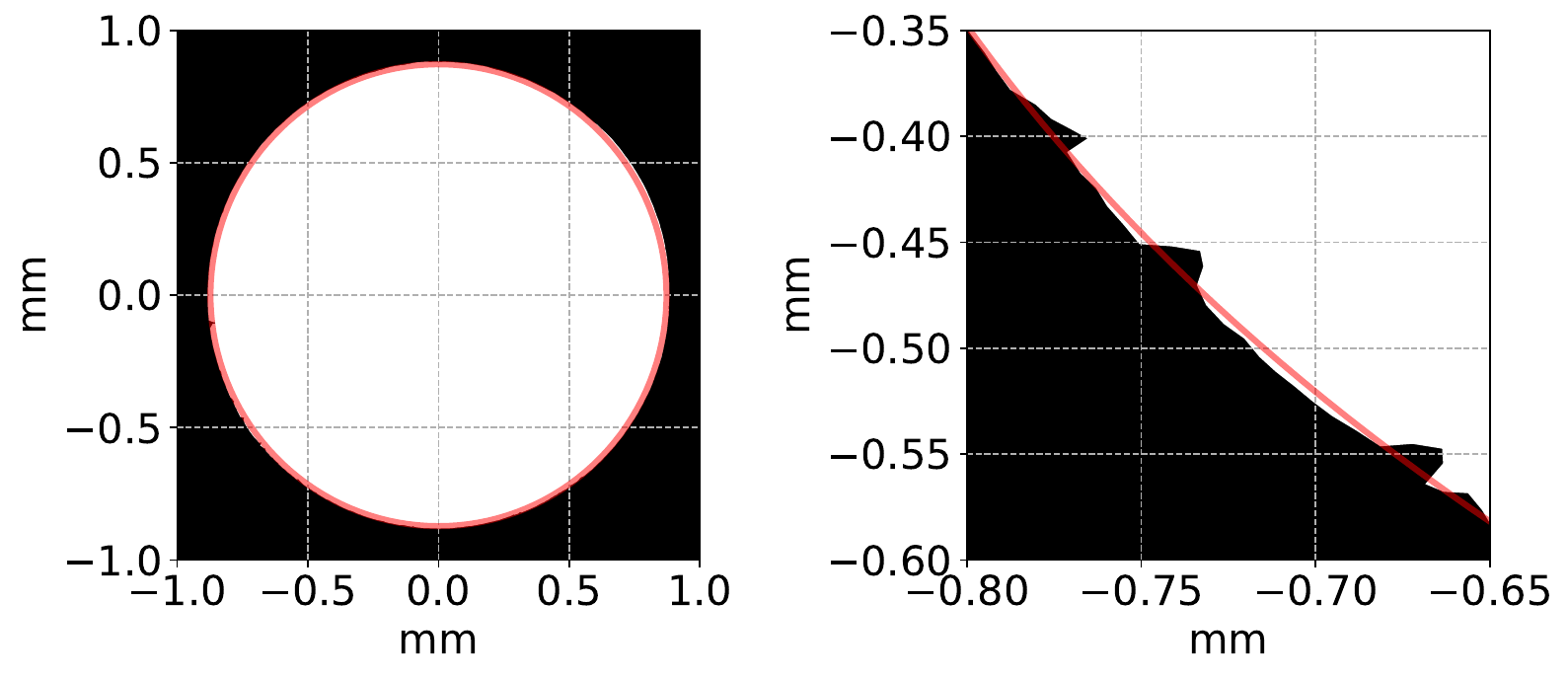}
            \newline
 		\centering\includegraphics[width=10cm]{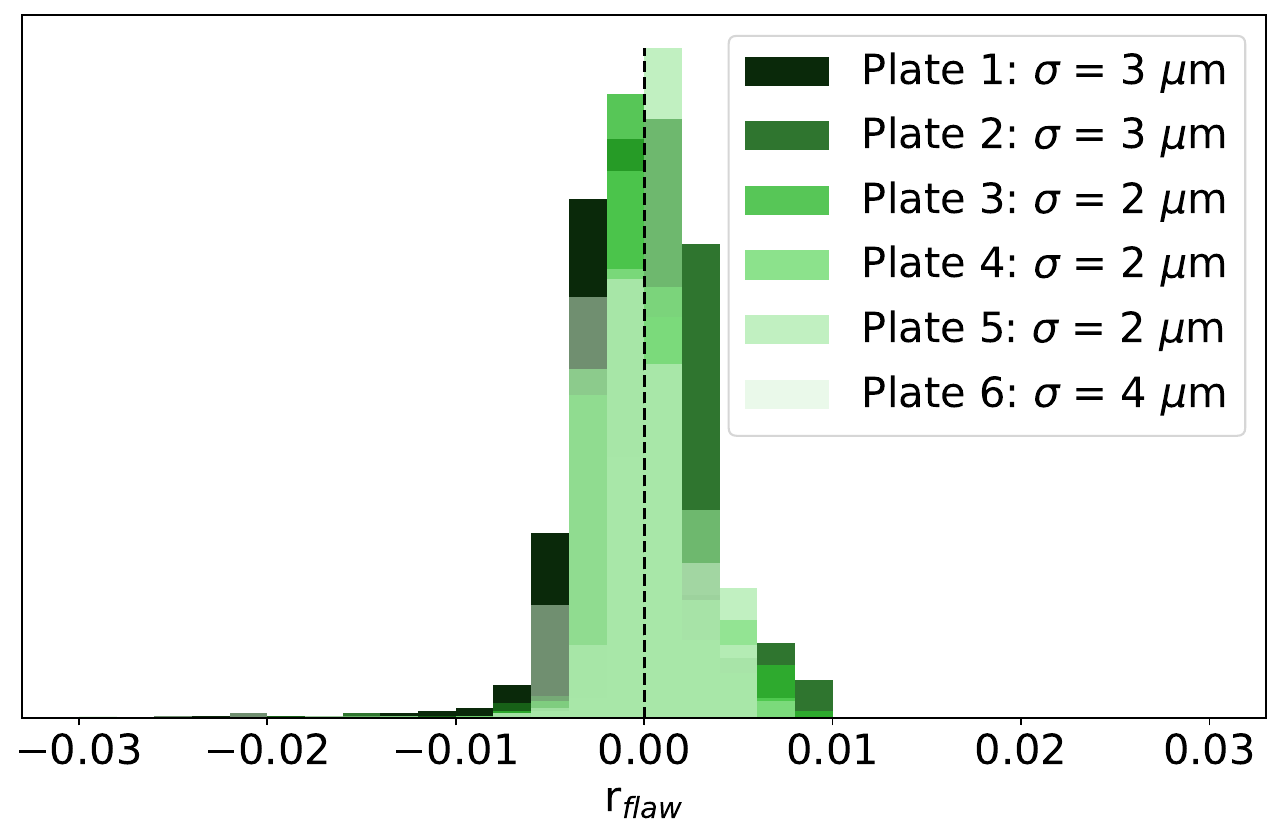}
	        \caption{\textit{Top panel}: Example of a reconstructed OMT plate using the point-cloud acquisition. We show the first 0.15\,mm plate, containing the circular waveguide, on the left. The metal plate is black and the cavity is white. The red circle is the best fit, obtained from the measured diameter. A close-up of the same plate is shown on the right, highlighting the details of the reconstructed profile, including the flaws and imperfections along the etched waveguide. \textit{Bottom panel}: distribution of the etching roughness, $r_{\mathrm{flaw}}$, for the six OMT plates containing the circular waveguide, quantified as the deviation of the radius given by the measured $x$ and $y$ point coordinates from the metrological best-fit radius. The distributions are centered in zero, with a tail at negative $r_{\mathrm{flaw}}$ due to the bumps along the etched profile, and with a $\mathrm{RMS}\simeq3\ \mu$m.}
     		\label{fig:flaws_distribution}
	   \end{figure}

\begin{figure}[ht]
            	\includegraphics[width=16cm]{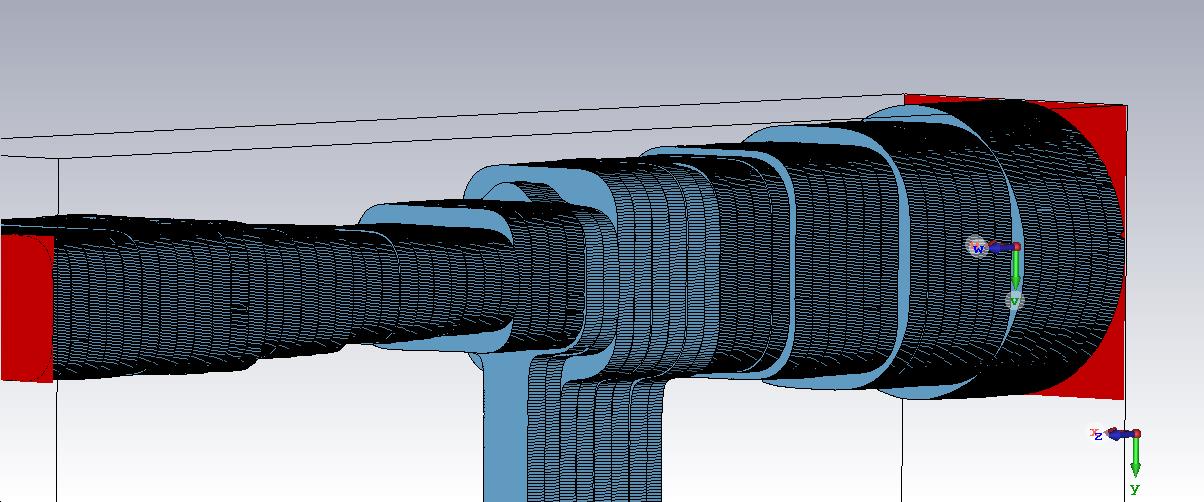}
            	\caption{Close-up on the 3D model of the manufactured OMT reconstructed from the metrological data.}
         		\label{fig:measured_OMT_em_design}
    	\end{figure}

        We performed a \textit{point-cloud acquisition} on each plate using our Werth metrological machine, which consists of the semi-automatic motion of the optical sensor along the OMT profile and the simultaneous measurement of the point coordinates, in the machine $x$-$y$ reference frame, with a fixed sampling (around 50\,pts/mm). The points are saved in a text file, thus allowing us to reconstruct the measured profile \textit{a posteriori}, as shown in the example in the top-panel of Fig.~\ref{fig:flaws_distribution}. We imported the datasets into CST Microwave Studio to reconstruct the 3D model of the measured OMT, as can be seen in Fig.~\ref{fig:measured_OMT_em_design}, which contains the manufacturing defects and systematic effects both in the geometry and position of the platelets.
        
        In the CST model we used a coarser mesh compared to the measurement resolution of 50 points per mm in order to maintain the total number of cells within $\sim$2$\times$10$^6$, which was compatible with the available computing resources.
        
        With the measured data we simulated three cases:
        \begin{enumerate}
            \item[A.] the OMT with the measured profile;
            \item[B.] an OMT in which we shifted all the measured profiles in the origin of the coordinate system to simulate the over-eroded, but aligned cavities;
            \item[C.] as in case B with a random misalignment in the cavity centers to study the sensitivity to a plate-to-plate misalignment induced during the mechanical assembly.
        \end{enumerate}

        Case C was particularly interesting as mechanical misalignment is a well-known cause of performance reduction in waveguide OMTs \cite{Navarrini_2011_test}. In our case, although we used two orthogonal grinded surfaces as support during the assembly process, some level of plate misalignment could still arise from the pin holes accuracy. 
    
        To study the sensitivity of our prototype to this effect, we added a $(\Delta x,\Delta y)=(\Delta r\cos\theta, \Delta\mathrm{r}\sin\theta)$ to the measured position of each plate, where $\theta$ is randomly distributed between $[0,2\pi]$ and $\Delta r$ is randomly sampled from a Gaussian distribution centered around $\mu=0$\,mm, with $\sigma$ equal to the mean pin radius deviation ($\sigma=0.019$\,mm). This represents the situation in which the OMT is assembled along the input waveguide axes (distribution centered in $\Delta r=0$\,mm) with a plate-to-plate misalignment.
        
         \subsubsection{Simulated electromagnetic performance}
            \label{Sim_of_metrol_OMT}
            
\begin{figure}[h!]
            	\includegraphics[width=16cm]{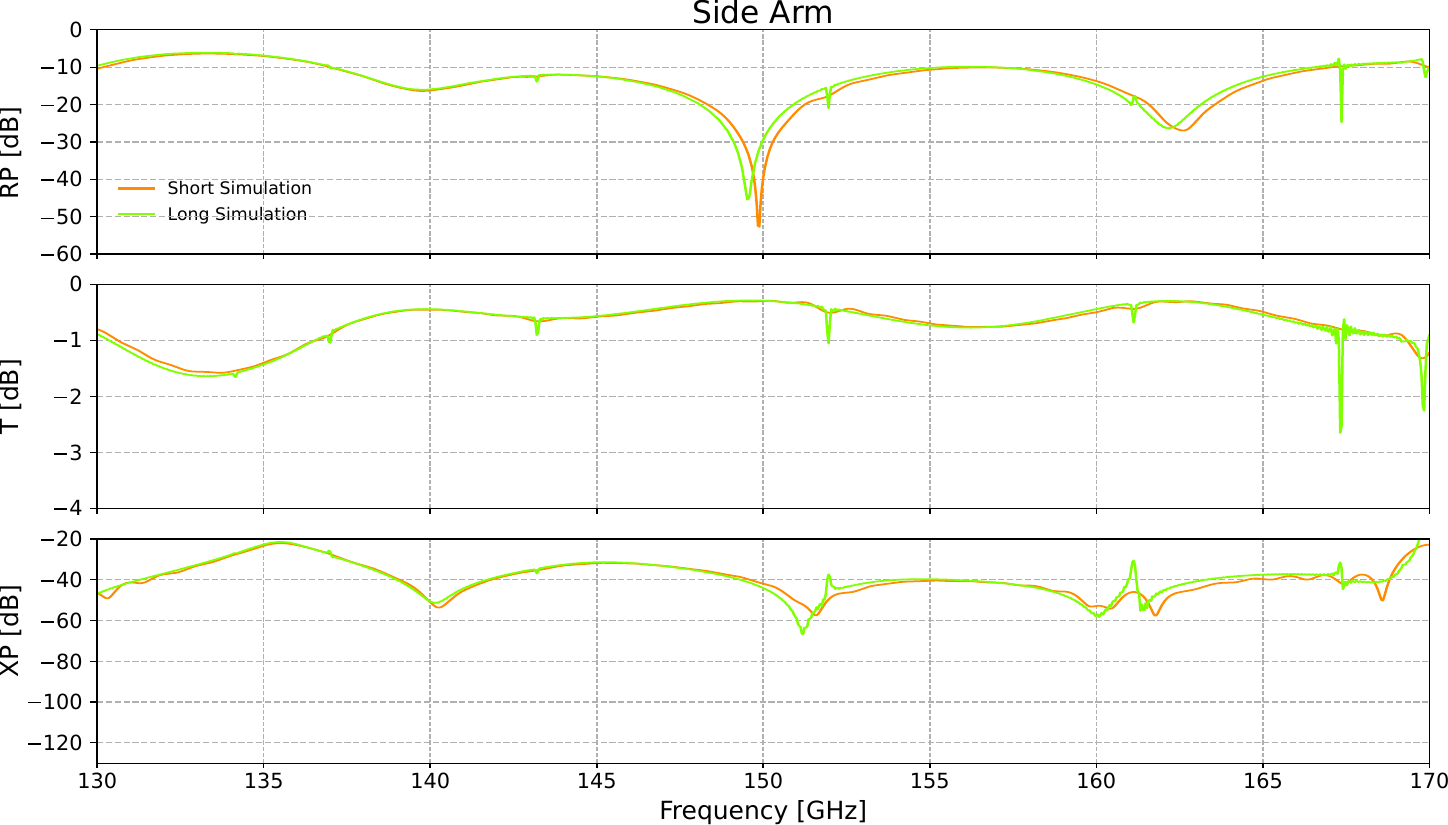}
            	\caption{{Comparison of simulated performance for the nominal case with different accuracy thresholds: \textit{Orange}: short simulation ($-30$\,dB accuracy); \textit{light green}: long simulation ($-40$\,dB accuracy).}}
         		\label{fig:long_short_comparison}
        	\end{figure}

        	{We performed CST time-domain simulations with an accuracy of $-30$\,dB, a threshold that allowed us to keep the computational time within reasonable limits (about 90 minutes per simulation). We also simulated the nominal configuration of the side arm with a higher accuracy $-40$\,dB (corresponding to about 8 hours of computation time) to check for possible effects not captured by the shorter simulations (see Fig.\,\ref{fig:long_short_comparison}). The comparison shows that the shorter simulation captures the main features in the electromagnetic response, apart from some sharp, narrow peaks that are present only in the longer simulation that do not affect the general results of our paper.}

\begin{figure}[ht]
            	\includegraphics[width=16cm]{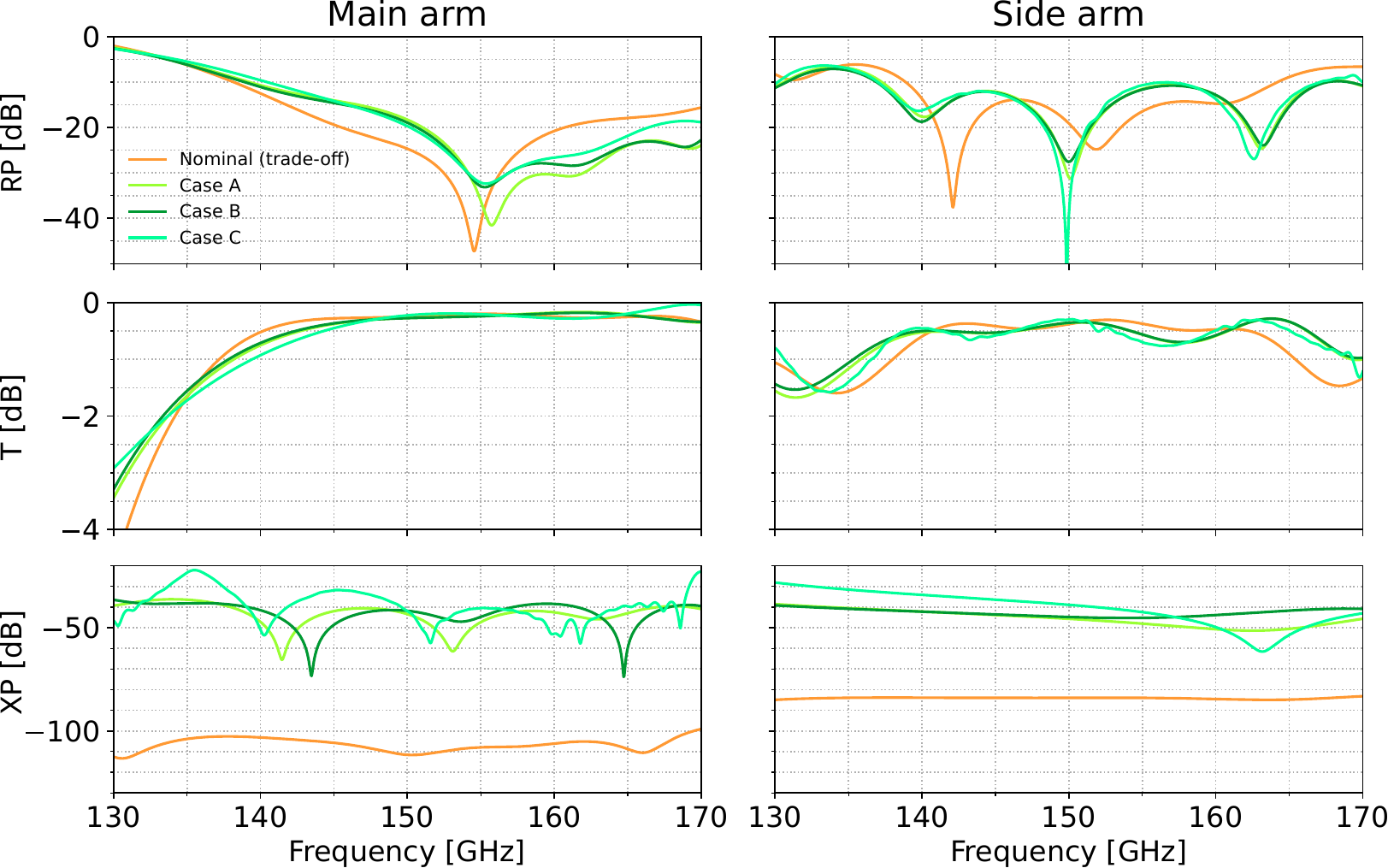}
            	\caption{Comparison of simulated performance. \textit{Orange}: nominal (trade-off) case; \textit{light green}: case A (profile with measured data); \textit{dark green}: case B (measured profile with aligned centers); \textit{light blue}: case C (case B with random misalignments. The OMT performance is mainly altered by the discontinuities of the over-etched profile.}
         		\label{fig:Sim_OMT_meas_vs_nom}
        	\end{figure}

        	Figure~\ref{fig:Sim_OMT_meas_vs_nom} compares the simulations of the measured OMT, the measured and aligned OMT, and the nominal (trade-off) design. As expected, the manufacturing systematic effects degrade the OMT's performance, particularly the {reflected power}, {cross-polarization and isolation}. However, the degradation is primarily attributed to the discontinuity between plates and the systematic effects in the geometry of the profiles, rather than their positions.
        	
        	Nonetheless, the {reflected power} remains below $-$10\,dB in the $140-160$\,GHz range and below $-$20\,dB at the center of the bandwidth. The transmission coefficient is approximately $-$0.5\,dB in the same range, and {both the cross-polarization and isolation remain} below $-$30\,dB across the entire bandwidth. The results also demonstrate that as long as the plates are properly aligned along the circular waveguide axes, the accuracy of the etched pin holes does not significantly affect the OMT performance.

\section{Electromagnetic measurements}\label{EMmeas}

In this section, we present the electromagnetic measurements that were carried out at the Physics Department of the University of Milano-Bicocca. In particular, we describe the instrumental setup used to measure the frequency response of the OMT prototype, we present the measured scattering parameters and discuss the transmission and {cross-polarization} results. 

    \subsection{Measurement setup and ``shorted'' method}\label{Setup}
    %what it is, how it is usually done, how we do it, and what we expect.
    %which vna and frequency extensions we used, cal. type, interface measurements with short and S21 only for bottom flange and results. OMT meas setup and say that for the isolation we tried both twist and rotation of the RX

\begin{figure}[ht]
        	\includegraphics[width=8cm]{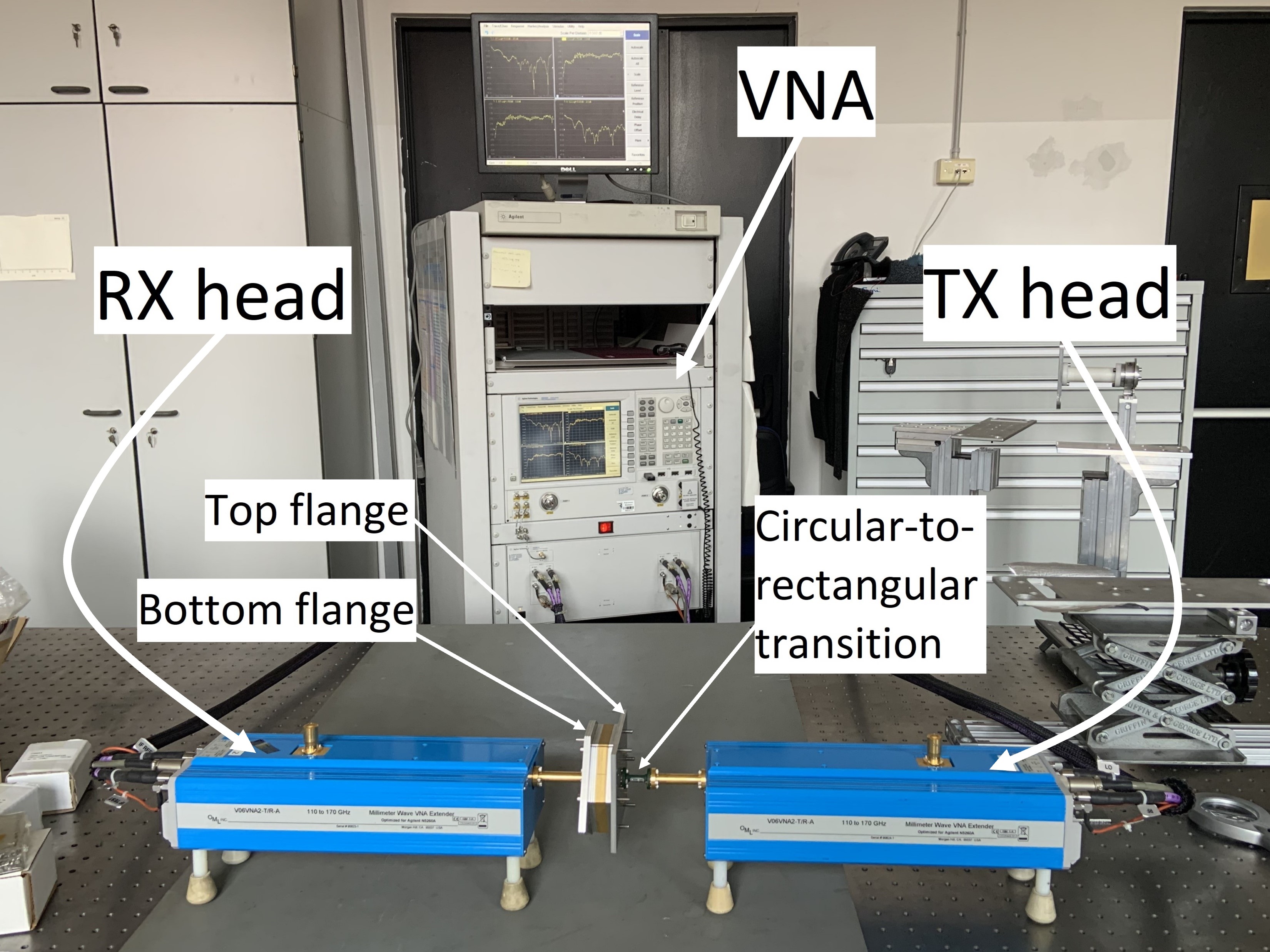}
        	\includegraphics[width=8cm]{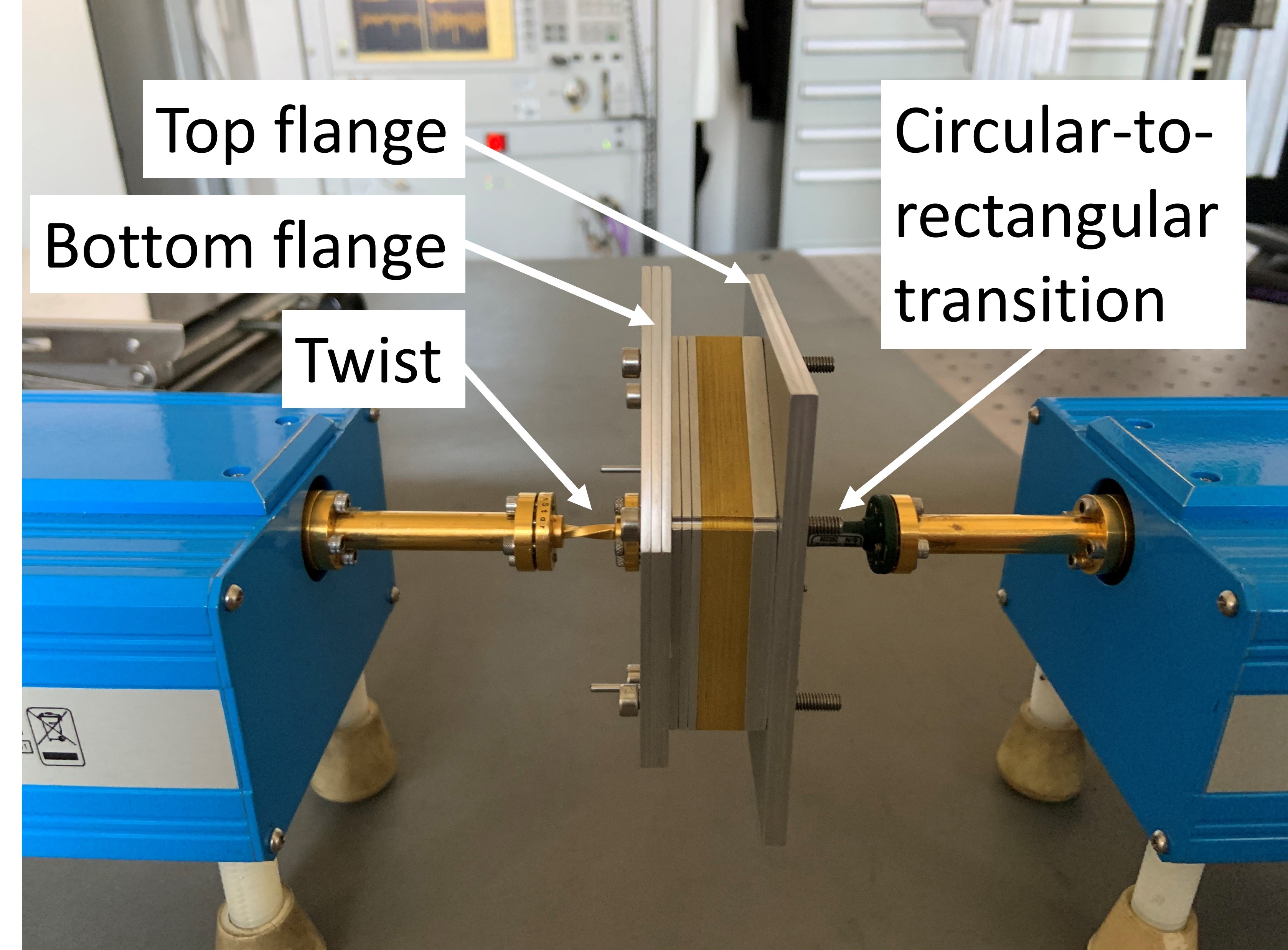}
        	\caption{Experimental setup. \textit{Left panel}: the transmission line. The OMT is connected to the TX frequency multiplier through a flange and a circular-to-rectangular transition on the right, and to the RX frequency multipliers on the left, through a second flange that simultaneously blocks the output port not under test. The rectangular waveguides of the TX and RX heads are coherently aligned and powered by the VNA in the back. \textit{Right panel}: the {cross-polarization} measurement. The measured output port is the one rotated by $90^{\circ}$ with respect to the incoming TX radiation and is connected to the RX head through a $90^{\circ}$ twist.}
     		\label{fig:setup}
	\end{figure}

    The OMT is connected to a vector network analyzer (VNA) by means of two frequency multipliers: a transmitter (TX head), that multiplies the low-frequency signal of the VNA ($10-20$\,GHz) to high-frequency, and a receiver (RX head) that down-converts the high-frequency signal leaving the OMT to low frequency back into the VNA. The setup is shown in Fig.~\ref{fig:setup}. We used two OML millimeter extensions (model V06VNA2-T/R-A) for full 2-port S-matrix characterization in the 110–170 GHz band and an Agilent PNA-X VNA (model N5246A). We calibrated the system using a 2-port TRL (Thru-Reflection-Line) calibration kit from OML
    
    The TX and RX heads interface with the OMT through a rectangular waveguide with a standard UG387/U-mod flange. However, none of the OMT top and bottom flanges can interface to a standard flange. We, therefore, used two additional flanges (top and bottom) that extend the OMT port with either a circular or an oval waveguide and that can be screwed to the OMT on one side and interface to the standard flange on the other. A circular-to-rectangular transition is screwed onto the top flange to interface with the TX head.

    Conventionally, the scattering parameters are measured ensuring impedance matching along the waveguide chain. During reflection measurements, the OMT input port is connected to the TX head, which transmits the signal and receives it back, while each OMT output port is terminated with a load that absorbs any transmitted radiation. During transmission and {cross-polarization} measurements, the input and one of the output ports are connected to the TX and RX heads, respectively, while the port not under test is terminated with a load.

    However, the two output ports of our OMT are too close to each other to simultaneously connect them both to a standard flange (load or extension), since their center-to-center distance is only 4.22\,mm. Therefore, we used a slightly different experimental setup: we fabricated the bottom flange with only one oval waveguide, so that it connects the port under test to the measurement chain (the RX head or a load) and, at the same time, the metal part blocks the other output port. The radiation coming to the port not under test is thus reflected, instead of absorbed, and creates stationary waves inside its arm. We call this method \textit{shorted}. % since it reminded us of a short circuit.
    
\begin{figure}[ht]
        	\includegraphics[width=16cm]{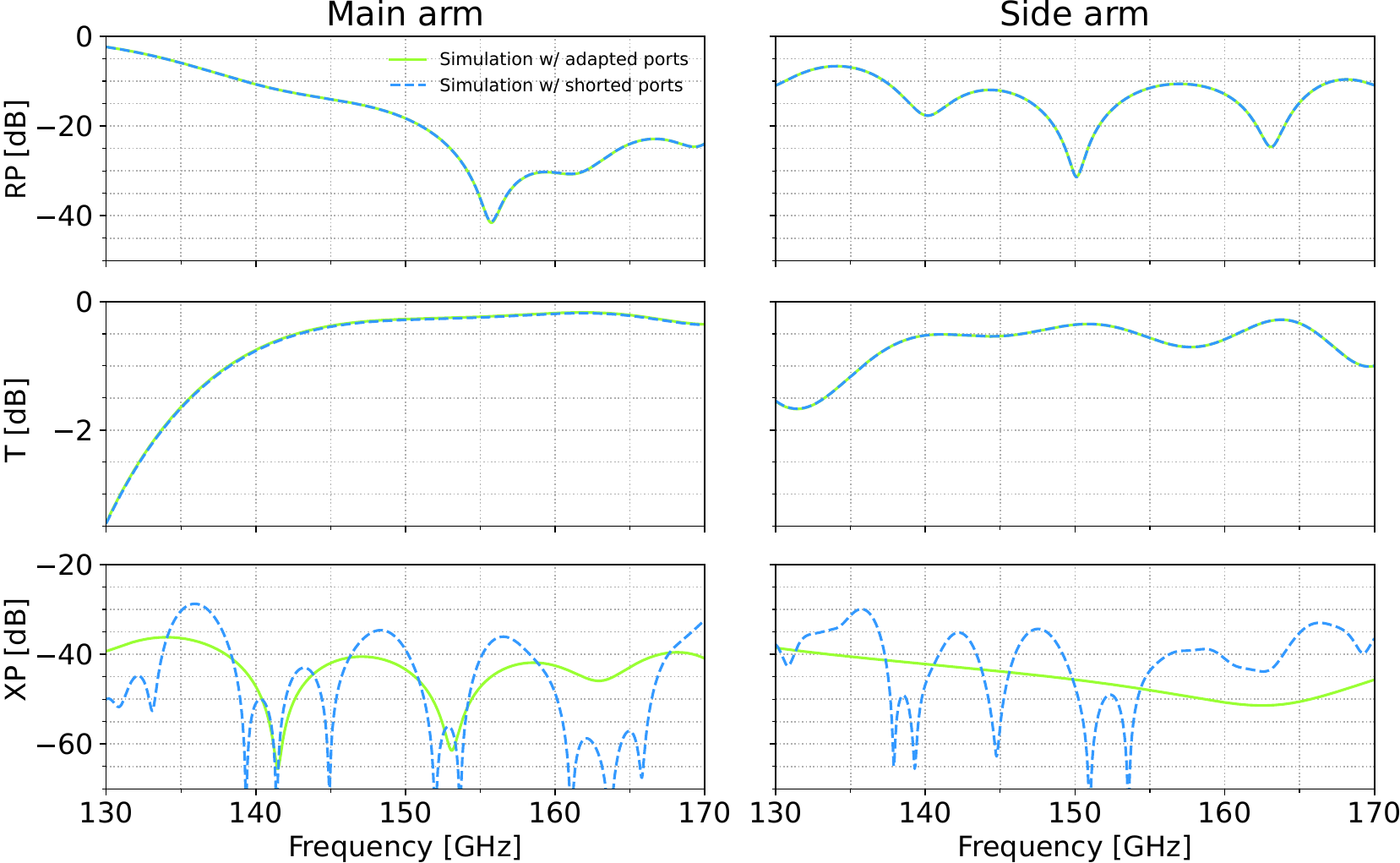}
        	\caption{The expected performance using our \textit{shorted} measurement setup compared with the true scattering parameters, namely the ones we would expect to measure if we properly adapted the OMT ports. The solid green curve is the metrologically measured OMT, whereas the dashed sky blue curve is the same OMT but with the port not under test blocked out by a metal plate (shorted).}
     		\label{fig:sim_port_adapted_vs_shorted}
	\end{figure}

    To compare the performance of the shorted method with the rigorous, adapted one, we simulated the measurement in CST with both methods. Figure~\ref{fig:sim_port_adapted_vs_shorted} compares the two simulated responses and shows that reflection and transmission curves essentially overlap. During {cross-polarization} measurements, instead, the stationary waves form in the arm that is aligned with the incoming polarization and, therefore, receive the majority of the signal. Part of the stationary signal is transmitted back into the arm under test and generates the resonances that are visible in the {cross-polarization} plot. Nevertheless, the average value sets at a level below $-30$\,dB. 
    
    {A fraction of the reflected signal, however, can propagate through the output port due to imperfect isolation. This fraction impacts the cross-polarization measurements, so that the measured XP represents an upper limit of the true OMT XP. Because the simulation results shown in Fig.~\ref{fig:sim_em_nominal} indicate that I and XP are similar, we can expect the measured XP to exceed the true one by $\gtrsim 3$\,dB}
	
    \subsection{Results}\label{Results}
\begin{figure}[ht]
        	\includegraphics[width=16cm]{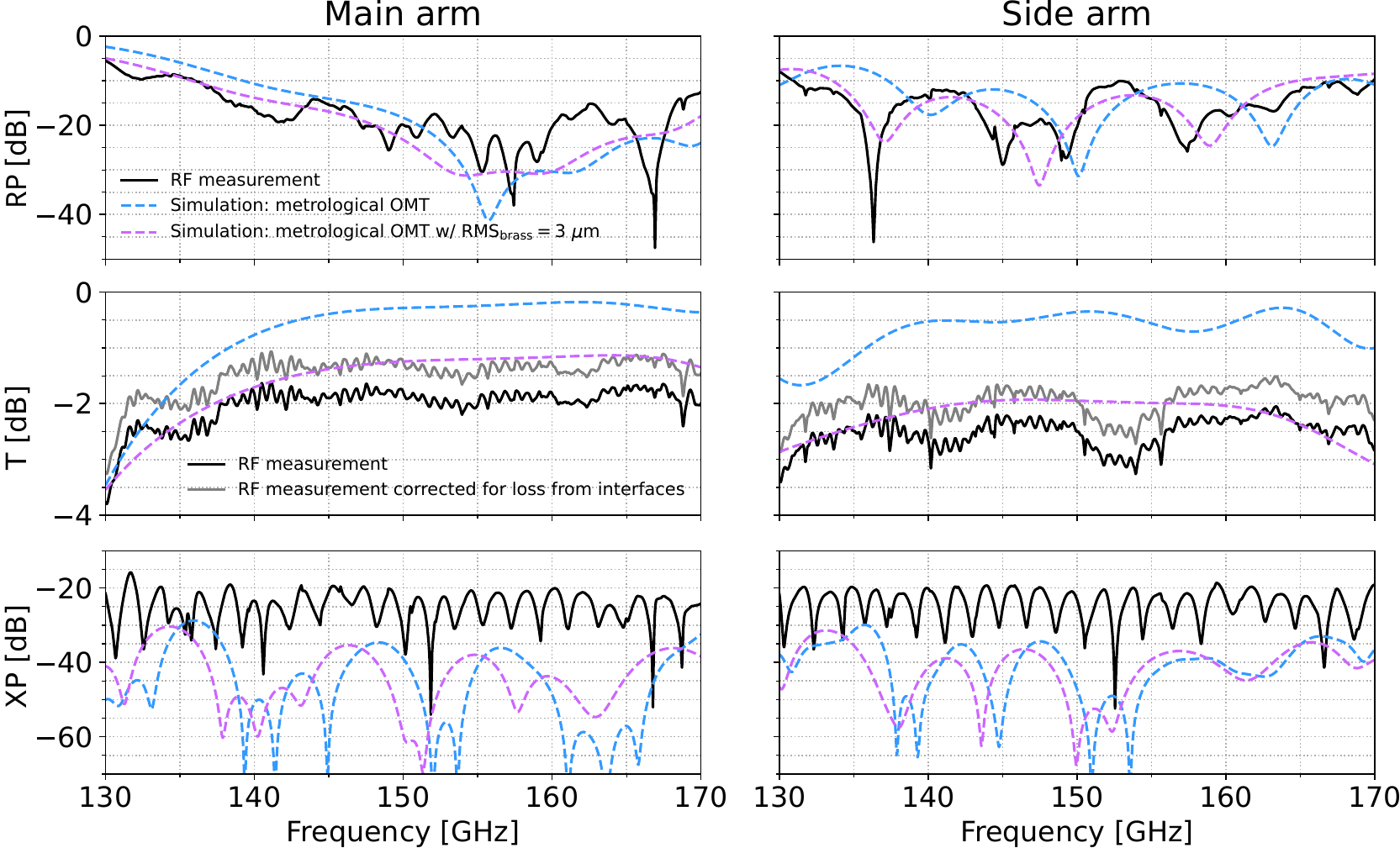}
        	\caption{\label{fig:meas_vs_sim}The measured performance, in solid black, compared to the expected, simulated one in dashed sky blue, namely, the metrologically measured OMT in the shorted configuration. The transmission measurement includes the OMT and the interfaces: the top and bottom flanges, and the transition. The estimated OMT transmission is shown in solid gray, where we subtracted the measured contribution of the interfaces from the black curve. The dashed purple line is the simulation of the metrologically measured OMT in the shorted configuration, with an additional parameter, the root mean square of the surface roughness, that accounts for the measured flaws along the etched profile, which we set to $\mathrm{RMS}_{\mathrm{brass}}=3\ \mu$m.}
	\end{figure}

    Figure~\ref{fig:meas_vs_sim} compares the measured and expected scattering parameters. The measured {reflected power} is below $-$10\,dB over the $140-160$\,GHz range and below $-$20\,dB at the center of the bandwidth for both arms. Overall, the curve matches the simulation, even though the resonances are slightly shifted in frequency, especially in the side arm.
    
    The measured transmission is around $-$2\,dB and $-$2.5\,dB for the main and side arm, respectively. The measurement refers to the complete waveguide chain including the OMT and its interfaces (top and bottom flanges plus the transition). To isolate the OMT transmission, we measured the {transmission (T)} of each interface and subtracted it from the measurement. Specifically, we measured the {reflected power} of the interface terminated with a short and estimated the {T} as half the average signal. For the bottom flange, we also measured its transmission and checked that it was compatible with the estimated {T}, which confirmed the reliability of our {transmission} measurement strategy. %\todo[size=tiny]{Shall we add the IL in an appendix?}
    
    The total contribution of the interfaces is around 0.5\,dB. Therefore, the estimated OMT transmission is around $-$1.5\,dB and $-$2\,dB for the two arms, which is more than 1\,dB larger than the simulation. The fact that the side arm, which is the longer arm in terms of electromagnetic path, has a lower transmission than the main arm suggests that the cause is related to an increase of resistive losses due to surface roughness.
    
    Indeed, surface roughness is a well-known cause of transmission losses in the millimeter range \cite{Gold_Helmreich_2012,Gold_Helmreich_2015,Gold_2017}. If the frequency is high enough that the skin depth falls below the surface roughness dimension, the imperfections cause extra {parasitic} currents that increase the resistive losses \cite{Gold_Helmreich_2012}. Because we downgraded the mesh of our reconstructed OMT model (see Section \ref{Sim_with_systematics}) our simulations do not properly account for the effects of profile imperfections. 
    
    CST allows the simulation of this effect by specifying a parameter, the root mean square (RMS) of the roughness, which is used to compute an effective conductivity and solve the modified field equations, following the gradient model introduced in Gold and Helmreich \cite{Gold_Helmreich_2012}. We performed the simulation of the reconstructed OMT assuming a value of RMS compatible with the one obtained from the metrological measurements, $\mathrm{RMS}_{\mathrm{brass}}=3\ \mu$m, as indicated by the distribution in the bottom panel of Fig.~\ref{fig:flaws_distribution}. 
    
    The simulation results (displayed in Fig.~\ref{fig:meas_vs_sim}) {show} that this effect can reproduce the average measured transmission, {confirming that other sources of losses, like contact resistance and/or plate oxidation, are negligible compared to surface roughness.} %Our simulation also show that roughness} also causes a slight shift in the frequency of the resonances in the side arm that matches the measurements.  %This confirms that the measurements can be explained by the surface roughness effect related to the systematic imperfections of the etched profile.

    {To reduce the surface roughness at these frequencies we are exploring alternative metal cutting methods, such as laser micro-machining, the results of which will be reported in a paper currently in preparation. Although the {transmission} obtained in the prototype presented here is too high for CMB polarization experiments, it may still be compatible with other applications in astrophysics and telecommunication.}
    
\begin{figure}[ht]
        \includegraphics[width=16cm]{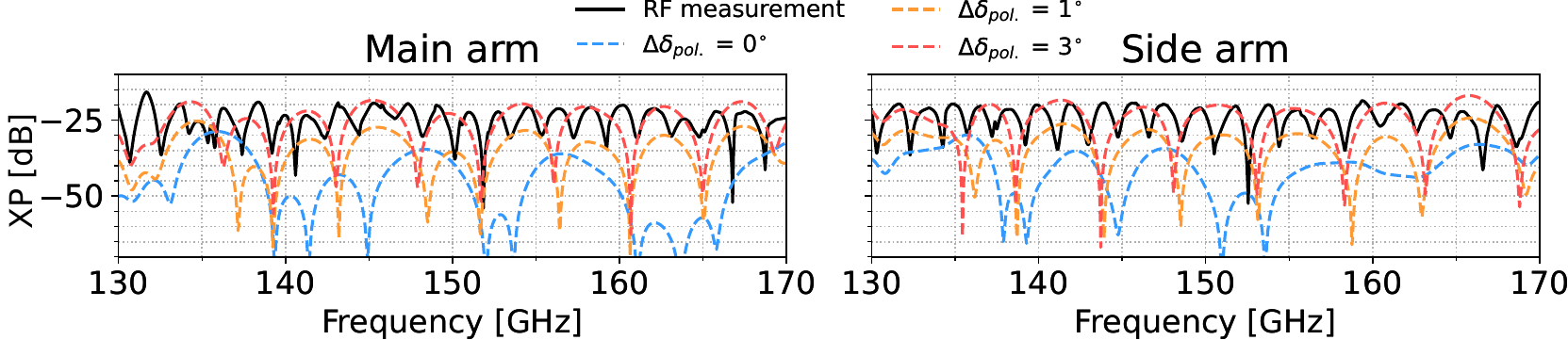}
        \caption{\label{fig:explain_isol_meas}Measured and simulated {cross-polarization} in the case of perfect horizontal or vertical polarization, $\Delta\delta_{pol}=0^{\circ}$, and with a one or three degrees rotation of the polarization angle, $\Delta\delta_{pol}=1^{\circ}$ and $\Delta\delta_{pol}=3^{\circ}$, respectively.}
    \end{figure}

    The measured {cross-polarization} is below $-$20\,dB along the full bandwidth, therefore around $-$10\,dB higher than the simulation. Further simulations suggest that this could be due to a slight polarization angle rotation within the measured setup. Figure~\ref{fig:explain_isol_meas} shows the comparison between measured and simulated {cross-polarization} assuming a $1^{\circ}$ and $3^{\circ}$ rotation and confirms that a $3^{\circ}$ rotation would be enough to explain the measured {cross-polarization}. We assessed the possible role of the $90^{\circ}$ twist that was placed between the output port and the RX head during the {cross-polarization} measurement by repeating the test without the twist and rotating the RX head around its side. Since we did not observe any difference in the measured {cross-polarization}, we excluded the twist as the cause, although we could not precisely determine where the rotation occurred along the remaining waveguide chain.
    
    However, we emphasize that even if the {measurement} represents the actual OMT {cross-polarization}, it remains comparable to the {cross-polarization} of planar OMTs currently used in CMB experiments \cite{McMahon_2012}.
    
    %Moreover, since our simulations in Section~\ref{Sim_with_systematics} indicate that the degradation in isolation arises from the measured profile, the performance could be improved in future prototypes by properly rescaling the OMT design to account for the over-erosion. \todo{check if it is the over-erosion or the plate-to-plate effect}

% 	\subsection{Discussion}\label{Discussion}
    We also checked whether a more conductive material, like aluminum (which is twice as conductive compared to brass), would lead to the same result in case of surface roughness. Therefore we simulated the metrologically measured OMT (case A) assuming completely made of aluminum with and without the same surface roughess considered for brass ($\mathrm{RMS}_{\mathrm{Al}}=\mathrm{RMS}_{\mathrm{brass}}=3\ \mu$m). 
    
\begin{figure}[ht]
        	\includegraphics[width=16cm]{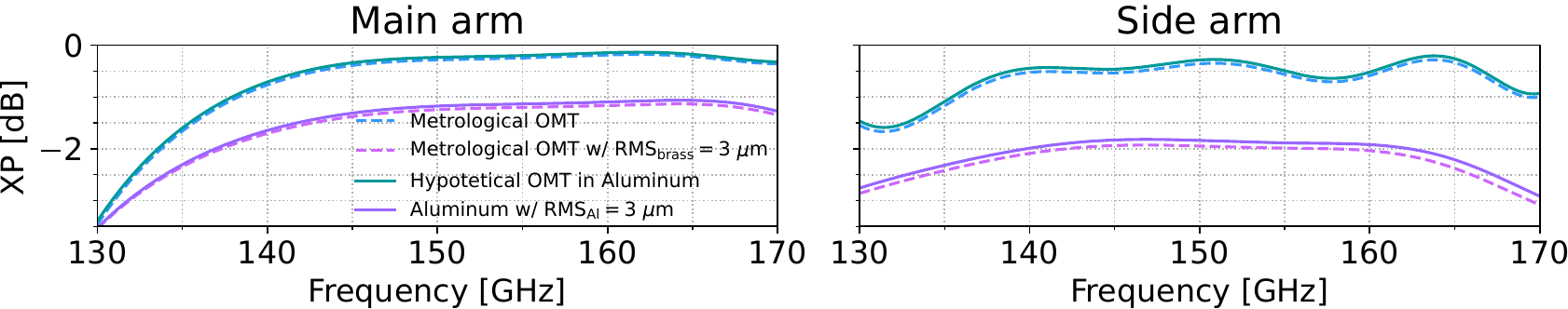}
        	\caption{\label{fig:alum_case}Simulated transmission for our metrologically measured OMT made of brass, with and without a root mean square surface roughness equal to the measured one, $\mathrm{RMS}_{\mathrm{brass}}=3\ \mu$m, in dashed purple and dashed blue, respectively. The curves are the same as shown in Fig.~\ref{fig:meas_vs_sim}. The same simulated transmission is shown with solid lines for a hypothetical, identical OMT, but completely made out of aluminum. Using aluminum instead of brass does not improve the loss in transmission if the manufacturing technique leaves imperfections along the OMT profile.}
        \end{figure}

    Figure~\ref{fig:alum_case} compares the simulated transmission for the two cases and confirms that the transmission loss is caused by the surface roughess alone, {independently} from the material. This result highlights a limit of chemical etching in all the applications in which resistive lossess, even of few dB, must be avoided.
	
%Discussions should be brief and focused. In some disciplines use of Discussion or `Conclusion' is interchangeable. It is not mandatory to use both. Some journals prefer a section `Results and Discussion' followed by a section `Conclusion'. Please refer to Journal-level guidance for any specific requirements. 

%~~~~~~~~~~~~~~~~~~~~~~~~~~~~~~~~~~~~~~~~~~~~~~~~~~~~~~~~~~~~~~~~~~~~~
\section{Conclusions}\label{Concl}

In this paper we presented the design, fabrication, and testing of the first published asymmetric waveguide OMT working in the D-band ($130-170$\,GHz) realized with chemically etched brass platelets that were subsequently aligned and mechanically clamped.

The aim of our study was to investigate fast, low-cost (less than 1 kEuro for this prototype), and scalable manufacturing techniques for producing large arrays of OMTs to be coupled to feed-horn antennas and KIDs arrays for measuring the CMB polarization. We focused on chemical etching in combination with the platelet manufacturing technique, which has already proven successful in producing corrugated feed-horn arrays with state-of-the-art performance up to $150$\,GHz. 

We performed both metrological and electromagnetic measurements on our prototype. The metrological tests confirmed a systematic tendency of chemical etching to over-erosion, which was noticed in previous works on feed-horn arrays. This effect can be corrected by properly rescaling the OMT profile during the design phase. Simulations based on the measured profile indicated that the over-erosion mainly degrades the {cross-polarization}, which increased to around $-30$\,dB, whereas the reflection losses and transmission remained compatible with simulations of the nominal OMT: {refleccted power} below $-10$\,dB across the $140-170$\,GHz range with a minimum of $-20/-35$\,dB at the center, and transmission above $-0.5$\,dB. The simulations also showed a small sensitivity to plate misalignment that could be induced by the pin holes accuracy during the mechanical assembly.

{To overcome mechanical constraints imposed by the distance between the output waveguides, we did not adapt the unmeasured port but shorted it with a metal plate. Then we compared the output measured power at the tested port with CST simulations of the same configuration. This method allows one to measure transmission, reflection, and a combination of isolation (I) and cross-polarization (XP). Because I and XP are at the same level the measurement provides a reasonable upper limit of XP. Although this method is less accurate than the rigorous, adapted procedure, it allows one to carry out measurements in cases where I and XP are approximately equal and it is complex to properly adapt the unused port.}

The measured frequency response showed {reflected power} compatible with simulations, but with resonances slightly shifted in frequency, a transmission of around $-1.5/-2$\,dB, compared to the expected $-0.5$\,dB, and {cross-polarization} around $-20$\,dB, instead of the expected $-30$\,dB. Further simulations have confirmed that both the transmission and the shift of the resonances can be explained by the surface roughness effect arising from defects along the etched profile ($\mathrm{RMS}\simeq 3\ \mu$m), which is a typical and unavoidable consequence of the chemical erosion.

Further simulations also showed that the increase in the {cross-polarization} could be explained by a slight rotation of the polarization angle within the measurement chain (around $3^{\circ}$), even though it was difficult to assess where precisely it could have occurred. Even if this was the actual response of the OMT, it would still be compatible with the {cross-polarization} of planar OMTs that are currently employed in CMB experiments.

In summary, our study illustrates that chemical etching has the potential to be a fast, low-cost, and scalable technique for producing waveguide OMTs with performance compatible with the state-of-the-art, as long as the OMT design is properly rescaled in the design phase to compensate for the systematic over-erosion of the plates. We have observed a degradation in transmission caused by the roughness of the etched profile, which could limit the applicability of chemical etching to OMTs above $100$\,GHz for CMB polarization measurements. We believe that chemical etching could still be suitable for lower-frequency OMTs, where the surface roughness should have a smaller impact or for other applications, where an insertion loss of few dB does not represent an issue

For high-frequency devices, we are currently investigating a alternative solutions, like laser micro-machining, that could still guarantee a fast and low-cost production, but with a smoother finish. Our results will be deferred to a dedicated paper.

\backmatter

\bmhead{Acknowledgements}

    This project has been carried out in the framework of the \textit{Kinetic Inductance Detector for Space} and \textit{Premiale} projects funded by the Agenzia Spaziale Italiana (ASI). The authors wish to thank the entire staff of the machine shop at the Physics Department of the University of Milan, for their contribution and key discussions about the realization and assembly process.

\bibliography{Ref}% common bib file
%% if required, the content of .bbl file can be included here once bbl is generated
%%\input sn-article.bbl

\end{document}